\documentclass[twocolumn,showpacs,preprintnumbers,amsmath,pre,prb]{revtex4-2}
\usepackage{graphicx}
\usepackage{dcolumn}
\usepackage{bm}
\usepackage{color}
\usepackage{hyperref}
\usepackage{tabularx}

\usepackage{graphicx}
\usepackage[normalem]{ulem}
\usepackage{color}
\usepackage{amsmath}
\usepackage{enumitem}

\newcommand\wordcount{
    \immediate\write18{texcount -sub=section \jobname.tex  | grep "Section" | sed -e 's/+.*//' | sed -n \thesection p > 'count.txt'}
(\input{count.txt}words)}

\begin{document}

\author{David Soriano}
\affiliation{Dipartimento di Ingegneria dell’Informazione, Università di Pisa, Via G. Caruso 16, 56122 Pisa, Italy}
\affiliation{Departamento de Física Aplicada, Universidad de Alicante, 03690 Alicante, Spain}
\author{Damiano Marian}
\affiliation{Dipartimento di Fisica ``E. Fermi", Università di Pisa, Largo Pontecorvo 3, 56127 Pisa, Italy}
\author{Prabhat Dubey}
\affiliation{Dipartimento di Ingegneria dell’Informazione, Università di Pisa, Via G. Caruso 16, 56122 Pisa, Italy}
\author{Gianluca Fiori}
\affiliation{Dipartimento di Ingegneria dell’Informazione, Università di Pisa, Via G. Caruso 16, 56122 Pisa, Italy}
\email{gianluca.fiori@unipi.it}


\title{Strain-induced valley transport in CrBr$_3$/WSe$_2$/CrBr$_3$ van der Waals heterostructure}

\begin{abstract}
Two dimensional magnetic materials are at the forefront of the next generation of spintronic devices. The possibility to interface them with other van der Waals materials such as transition metal dichalcogenides has opened new possibilities for the observation of new and exiting physical phenomena. Here, we present a proof-of-concept valleytronic device based on CrBr$_3$-encapsulated WSe$_2$ showing an unprecedented valley splitting of $\sim 100$ meV under compressive strain of the WSe$_2$, able to be tuned by the relative magnetization of the encapsulating layers. Multiscale transport simulations performed on this device show a spin-valley current with a polarization higher than 80$\%$ than is maintained in a range of $\sim$~0.3~V gate voltage in a field-effect transistor configuration. The impact of the stacking configuration on the valley splitting is also evaluated.  
\end{abstract}

\maketitle


\section{Introduction}
The discovery of unprecedented and long-sought physical phenomena at the two-dimensional (2D) limit is opening new routes towards the design of novel ultrathin electronic devices. The combination of 2D materials with different electronic properties to form van der Waals heterostructures has made  possible the observation of new electronic, magnetic, and topological phases {\it via} proximity effects and non-trivial stacking configurations\cite{CaoJarillo2018,SongRoche2018,KarpiakDash2019,ZhongXu2020,KezilebiekeLiljeroth2020,LyonsTartakovskii2020}. In this regard, the recent advances on the synthesis and fabrication of 2D magnetic materials such as chromium trihalides (CrX$_3$), Cr$_2$Ge$_2$Te$_6$, and Fe$_3$GeTe$_2$ have become an important contribution for the realization of 2D-based proof-of-concept devices \cite{GongZhang2017,HuangXu2017,ChenGao2019,OchMattevi2021,GibertiniNovoselov2019, SorianoRossier2020}. 

The implementation of 2D magnetic materials in van der Waals heterostructures has led to the experimental observation of an unprecedented tunnel magnetoresistance in Graphene/CrI$_3$/Graphene \cite{KleinJarillo2018,GhazaryanMisra2018,WangMorpurgo2018}, topological superconducting phases in NbSe$_2$/CrBr$_3$ \cite{KezilebiekeLiljeroth2020} and NbSe$_2$/Cr$_2$Ge$_2$Te$_6$ \cite{AiDong2021} ferromagnetic Josephson junctions, and Néel-type skyrmions in WSe$_2$/Fe$_3$GeTe$_2$ \cite{WuKang2020} {\it via} proximity effects. However, the observation of a strong valley polarization (or valley Zeeman splitting) due to the combination of magnetism and spin-orbit coupling in van der Waals heterostructures combining a 2D ferromagnet (FM) with a transition metal dichalcogenide (TMD) is still elusive. The values reported so far are limited to 1-12 meV in CrI$_3$/WSe$_2$ \cite{ZhongXu2017,ZhangLiu2019,GeZhang2022,ZollnerFabian2023} and CrBr$_3$/MoSe$_2$ \cite{LyonsTartakovskii2020,CiorciaroImamoglu2020,GeZhang2022,ChoiCrooker2023} heterostructures, limiting their use in future valleytronic devices.    

\begin{figure}[th!]
\centerline{\includegraphics[width=\columnwidth]{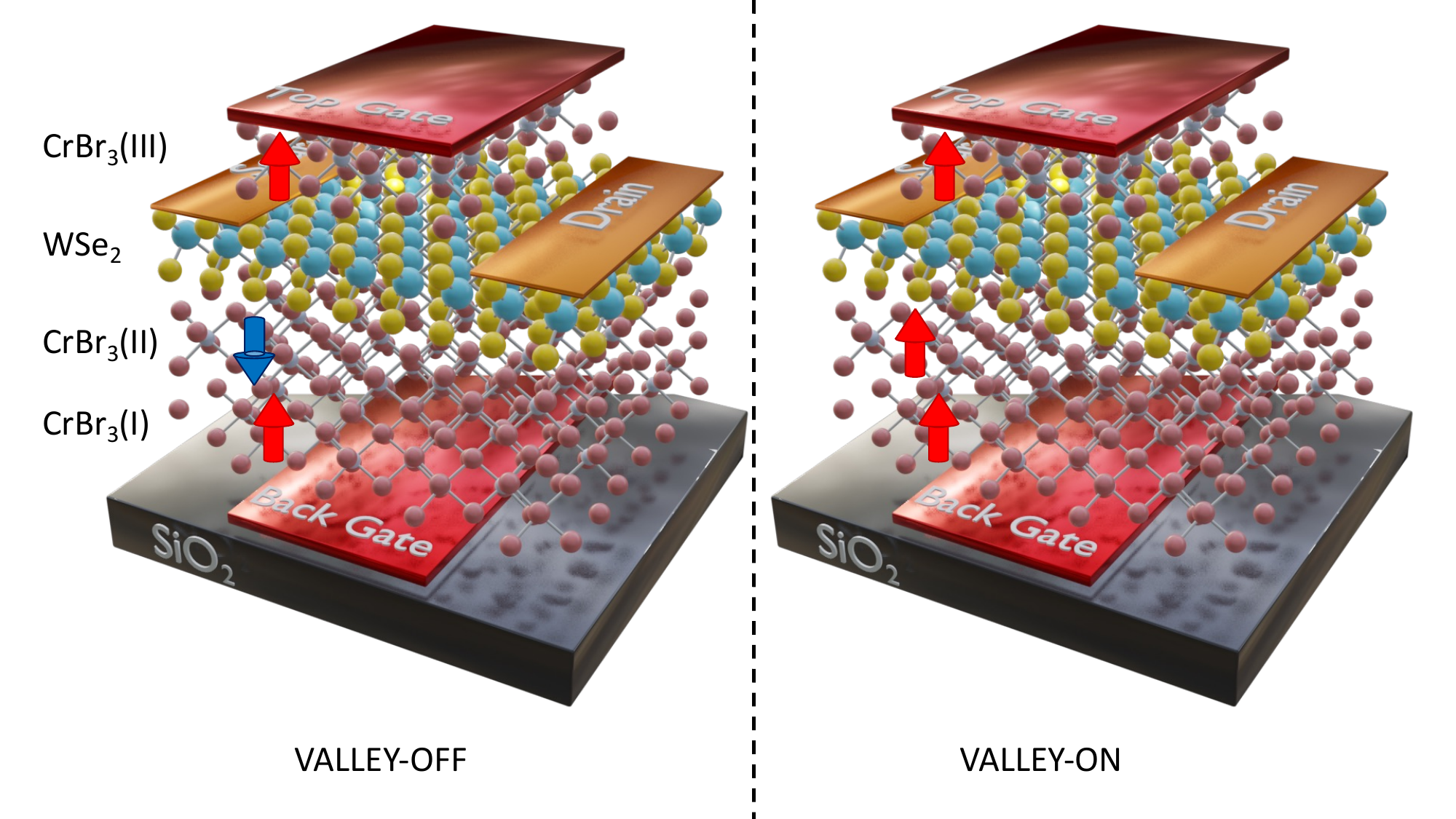}}
\caption{Schematics of the proposed proof-of-concept valleytronic field-effect transistor.}
\label{fig_device}
\end{figure}

In this article, we report a giant valley splitting of 100~meV in CrBr$_3$/WSe$_2$/CrBr$_3$ van der Waals trilayer heterostructure under compressive strain. The valley splitting takes place in the conduction band of WSe$_2$ due to the hybridization with the spin-polarized bands of the CrBr$_3$ layers. The valley splitting can be controlled through the relative orientation of the magnetization in both CrBr$_3$ layers, leading to K, K' , or 0 valley-polarized device. Based on that, we propose a proof-of-concept valleytronic field-effect transistor as the one depicted in Fig. \ref{fig_device}. The top magnetic layer CrBr$_3$(III) is pinned, while the bottom ones, CrBr$_3$(I) and CrBr$_3$(II), are free magnetic layers which allows to electrically control the magnetization and to turn ON/OFF the valley-polarization by a perpendicular electric field. In this proof-of-concept device, we assume that the bottom CrBr$_3$ bilayer is grown with a monoclinic stacking which results in an antiferromagnetic (AFM) interlayer ordering. By applying an external vertical electric field, it is possible to switch the interlayer exchange between layers I and II from AFM to FM \cite{JiangShan2018,JiangMak2018,HuangXu2018}, modifying the net magnetization affecting the TMD, from 0 (OFF) to a finite value (ON). Although exfoliated samples of bilayer CrBr$_3$ show a rhombohedral stacking with ferromagnetic interlayer magnetism, recent advances on molecular beam epitaxy (MBE) has made possible the growth of bilayer CrBr$_3$ samples with different stacking configurations showing antiferromagnetic interlayer exchange coupling \cite{ChenGao2019}. We assume that the device at source and drain is contacted with In metal, which ensures good electrons injection in the conduction band of WSe$_2$ \cite{Liu2013}.

The manuscript is organized as follow, In section II we show the band structure of CrBr$_3$/WSe$_2$/CrBr$_3$ trilayer heterostructure for different magnetic configurations using first-principles calculations. This is the region of the device where proximity effects play a key role in the development of the spin-valley splitting (the case with 4 layers is included in Appendix \ref{AppA}). We also show the band structure for different stacking configurations to evaluate its impact on the valley splitting. In section III, we propose an easy magnetic impurity model that describes the physics taking place in the heterostructure. In section IV, we show the performance of our proposed valleytronic device in realistic multiscale transport simulations, and in section V, we summarize the results.   

\section{First-principles calculations}
\begin{figure}[h!!!]
\centerline{\includegraphics[width=0.95\columnwidth]{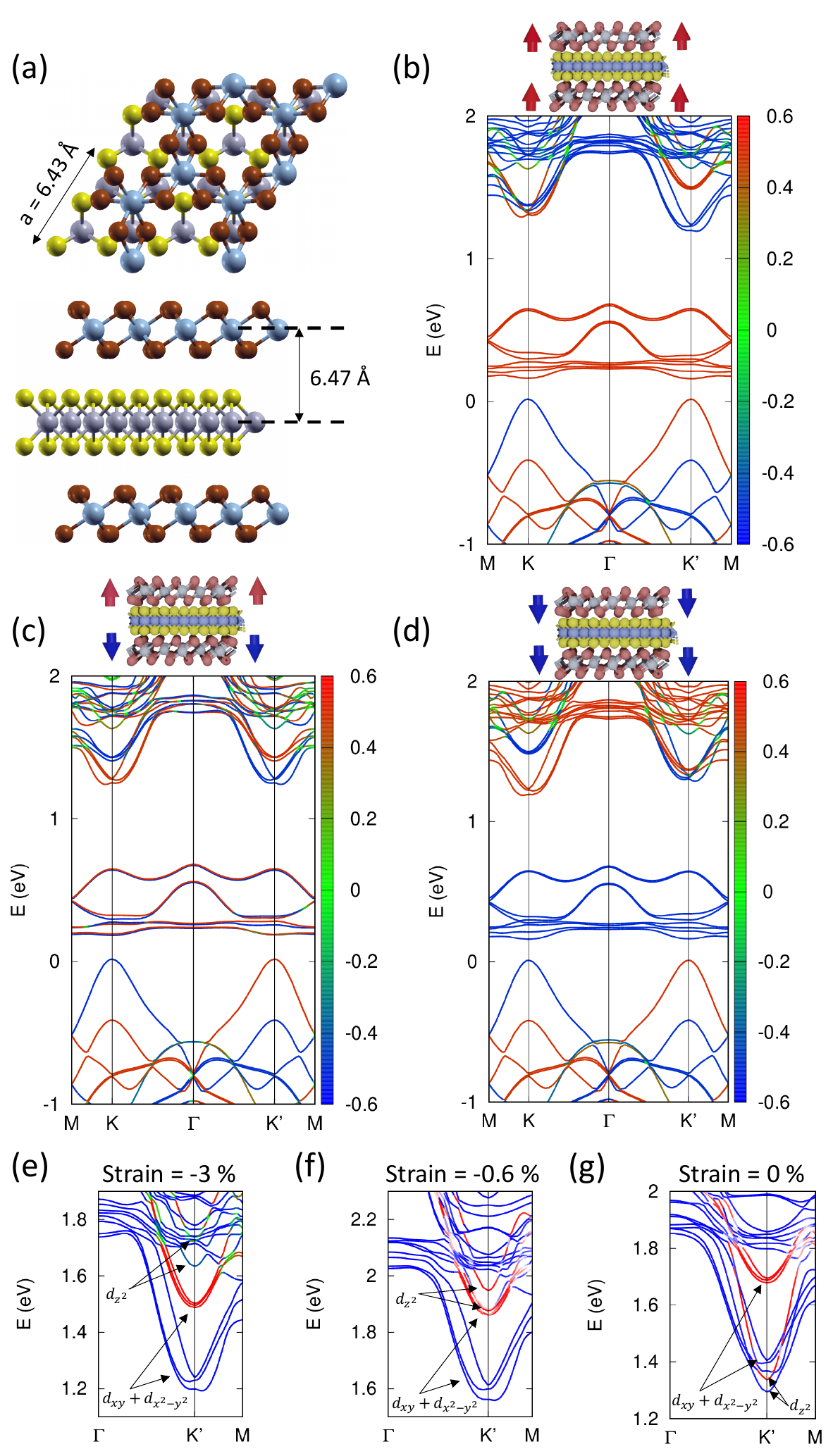}}
\caption{(a) Details of the rhombohedral stacking configuration of a 2x2 supercell of  CrBr$_3$/WSe$_2$/CrBr$_3$ trilayer heterostructure. Gray and yellow spheres correspond to W and Se atoms, respectively. Blue and brown spheres correspond to Cr and Br atoms, respectively. The unit cell size used in the calculations is indicated in the top panel with an arrow. The W atoms appear in the hollow sites formed by the Cr atoms. First-principles band structures for three different magnetic orientations: (b) CrBr$_3$ layers are co-polarized up, (c) CrBr$_3$ layers are counter-polarized, and (d) CrBr$_3$ layers are co-polarized down. Insets illustrate the magnetic configurations of the CrBr$_3$ layers in each case. Panels (e-f) show a zoom of the conduction bands around the K' point for three different strain values in WSe$_2$ layer in the van der Waals heterostructure. The Fermi level is set to E = 0. Color bar makes reference to the expectation value of the $z$-component of the spin $\langle S_z \rangle.$  }
\label{fig_DFT}
\end{figure}

To reduce the dimension and complexity of the calculations, we have carried out first-principles calculations on the CrBr$_3$(II)/WSe$_2$/CrBr$_3$(III) trilayer heterostructure with rhombohedral stacking between the layers as shown in Figure \ref{fig_DFT}(a), with the W (gray) atoms lying in the hollow and bridge positions of the encapsulating CrBr$_3$. We label this stacking R$_W$. The unit cell is built by combining a 2x2 supercell of WSe$_2$ with a single unit cell of the encapsulating CrBr$_3$ layers. To make them commensurate, the WSe$_2$ layer has been compressed 3\% with respect to the standard size, while the CrBr$_3$ layers have been stretched 0.8\% with respect to the standard lattice parameter. This strain, needed to commensurate the cells and to perform feasible simulations, significantly alter the electronic structure of WSe$_2$ moving the empty $d_{xy}$ and $d_{x^2-y^2}$ bands below the $d_{z^2}$ conduction bands, as shown in Figure \ref{fig_DFT}(e-g)\cite{DENG201844,SohierVerstraete2023}. These bands, i.e. $d_{xy}$ and $d_{x^2-y^2}$, hybridize stronger with the empty $t_{2g}$ CrBr$_3$ bands than the $d_{z^2}$ leading to an unprecedented valley splitting. Therefore, strain is key for the observation of this phenomenon. The band structures have been calculated using the Quantum-Espresso \emph{ab initio} package \cite{QE}. For the self-consistency, we have used fully relativistic projector augmented wave (PAW) pseudopotentials within the local density approximation (LDA), and local Hubbard (U~=~4~eV) and exchange (J~=~0.6~eV) corrections for the $3d$ orbitals of the Cr atoms, based on previous constrained random phase approximation (cRPA) calculations \cite{SorianoRosner2021}. The convergence threshold for the self-consistency has been set to $10^{-6}$ eV, and a 8x8x1 $\Gamma$-centered k-point grid have been used for all the cases studied. For the structure relaxation, we have included the Grimme-D2 van der Waals correction and we have let the forces relax until all the components for all the atoms in the trilayer heterostructure are smaller than $10^{-3}$ Ry/\AA.   

In Fig\ref{fig_DFT}(b-d), we show the band structures of CrBr$_3$/WSe$_2$/CrBr$_3$ heterostructure for three different magnetic configurations of the CrBr$_3$ encapsulating layers. When both CrBr$_3$ layers are co-polarized up(down) (Fig.\ref{fig_DFT}(b,d)), the down(up)-spin $t_{2g} \equiv \{\rm{d_{xy},d_{xz},d_{yz}}\}$ bands from the spin-polarized Cr atoms cross above the bottom of the conduction bands of WSe$_2$ at $\sim 1.8$ eV. These $d$-bands hybridize with the down(up)-spin $d_{xy}$ and $d_{x^2-y^2}$ bands of the WSe$_2$ layer breaking the spin-valley symmetry and leading to a valley splitting of 100 meV. It is very important to highlight here that band alignment is mandatory to observe this large spin-valley splitting. This occurs in the case of CrBr$_3$ and WSe$_2$, but it is not observed when replacing CrBr$_3$ with CrI$_3$, or WSe$_2$ by MoSe$_2$ (see Appendix \ref{AppB}).  When the magnetic layers become counter-polarized, both spin-up and spin-down $t_{2g}$-bands of the Cr atom hybridize with the $d_{xy}$ and $d_{x^2-y^2}$ bands of WSe$_2$, restoring the valley degeneracy as in Fig.\ref{fig_DFT}(c). 

In Figs.\ref{fig_DFT}(e,f,g) we report a zoom of the conduction bands of the WSe$_2$ in the heterostructure when different percentage of compressive strain is considered (-3\%, -0.6\%, and 0\%). As observed in the previous paragraph, strain is needed to commensurate the cells but it is also essential to move the $d_{xy}$ and $d_{x^2-y^2}$ bands (the ones that hybridize most with $t_{2g}$ CrBr$_3$ bands) below the $d_{z^2}$. We notice, however, that already for a compressive strain of -0.6\% the valley splitting is present. It is important to note that the valley splitting obtained for this heterostructure is not observed if the magnetic or TMD layers are replaced (see Appendix \ref{AppB}). 

\section{Stacking dependent valley splitting in C\lowercase{r}B\lowercase{r}$_3$/WS\lowercase{e}$_2$/C\lowercase{r}B\lowercase{r}$_3$ heterostructure}

\begin{figure}[t]
\centerline{\includegraphics[width=\columnwidth]{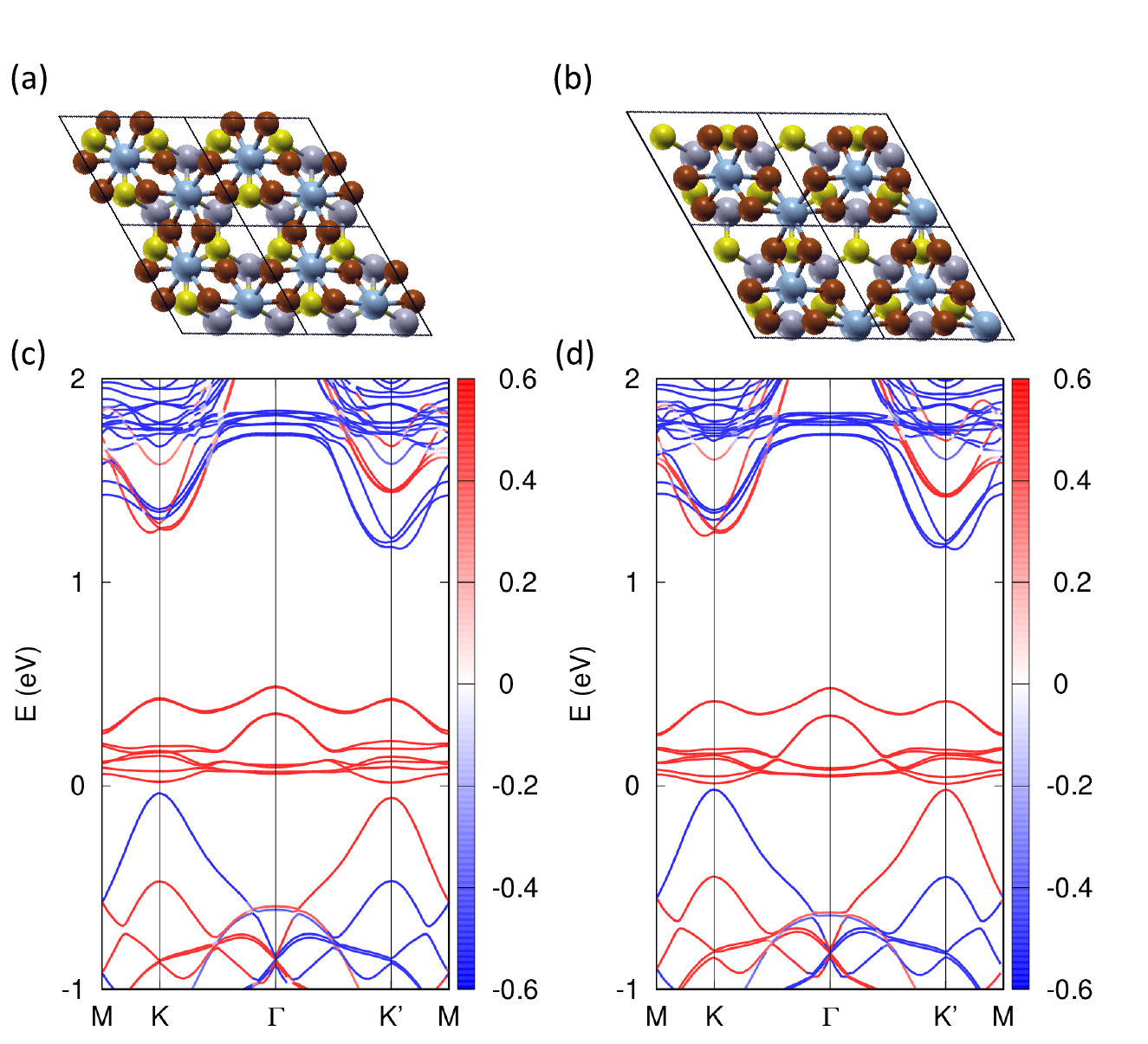}}
\caption{Band structure of CrBr$_3$/WSe$_2$/CrBr$_3$ heterostructure for the other two stacking configurations studied (b) M, and (c) R$_{Se}$. The structural details are shown in the top panel.}
\label{fig_SI_stack}
\end{figure}

Since stacking can affect the spin-valley splitting, as previously reported \cite{ZhangLiu2019,GeZhang2022,ZollnerFabian2023}, we have calculated the band structure for other two different stacking configurations. In Figs.\ref{fig_SI_stack} (c) and (d), we show the electronic band structure for the case in which the stacking configuration between layers is monoclinic (M) and the case in which the stacking is rhombohedral, like the one reported in the previous Section, but the hollow sites are occupied by Se atoms (R$_{Se}$). The structural details are reported in Figs.\ref{fig_SI_stack} (a) and (b) for the two stackings respectively. 
We compare these two stackings with the one (R$_W$) reported in the previous Section, i.e. rhombohedral with the W atoms appearing in the hollow sites of the Cr hexagonal lattices, in terms of ground-state energy, valley splitting in the conduction and valence band of the WSe$_2$. In this stacking (R$_W$) the valley splitting is the higher of the three cases studied (101 meV), while the valley splitting in the valence band is lower than 1 eV and the energy is $97$ meV above the stacking (M), which is the one with the lowest energy. In the monoclinic stacking the valley splitting in the conduction band decreases to 79 meV, however, this is the only case where strong valley splitting is observed in the valence band (21 meV). Finally, we discuss stacking (R$_{Se}$). This is the stacking with the higher energy. It gives rise to the smaller valley splitting in the conduction band (67 meV). Similar to the R$_W$ stacking, the valley splitting in the valence bands is lower than 1 meV.
These results demonstrate that the change in the stacking affects the hybridization of the $d$-orbitals in different layers (which is the main mechanism of valley polarization) leading to different values of the valley splitting. These results are summurized in Table \ref{tab_SI_sum}. More importantly, in all cases a valley splitting is observed higher than 60 meV making this heterostructure a promising candidate for future valleytronic proof-of-concept devices.

\begin{table}
    \caption{Energy difference with respect to the stacking with the lower energy ($\Delta E_{st} = E_{st} - E_{st_0}$) and valley splitting in conduction ($\Delta_{V_c}$) and valence ($\Delta_{V_v}$) bands of CrBr$_3$/WSe$_2$/CrBr$_3$ heterostructure for three different stacking configurations.}
    \centering
    \begin{tabularx}{\columnwidth}{X X X X}
        \hline
         & $\Delta E_{st}$ (meV) & $\Delta_{V_c}$ (meV) & $\Delta_{V_v}$ (meV) \\ 
         \hline
         R$_W$    & 97  & 101 & $<$1\\
         M        &  0  & 79  & 21 \\
         R$_{Se}$ & 343 & 67  & 1 \\
         \hline
    \end{tabularx}
    \label{tab_SI_sum}
\end{table}

\section{Model and theoretical description} 
\begin{figure}[t]
\centerline{\includegraphics[width=0.9\columnwidth]{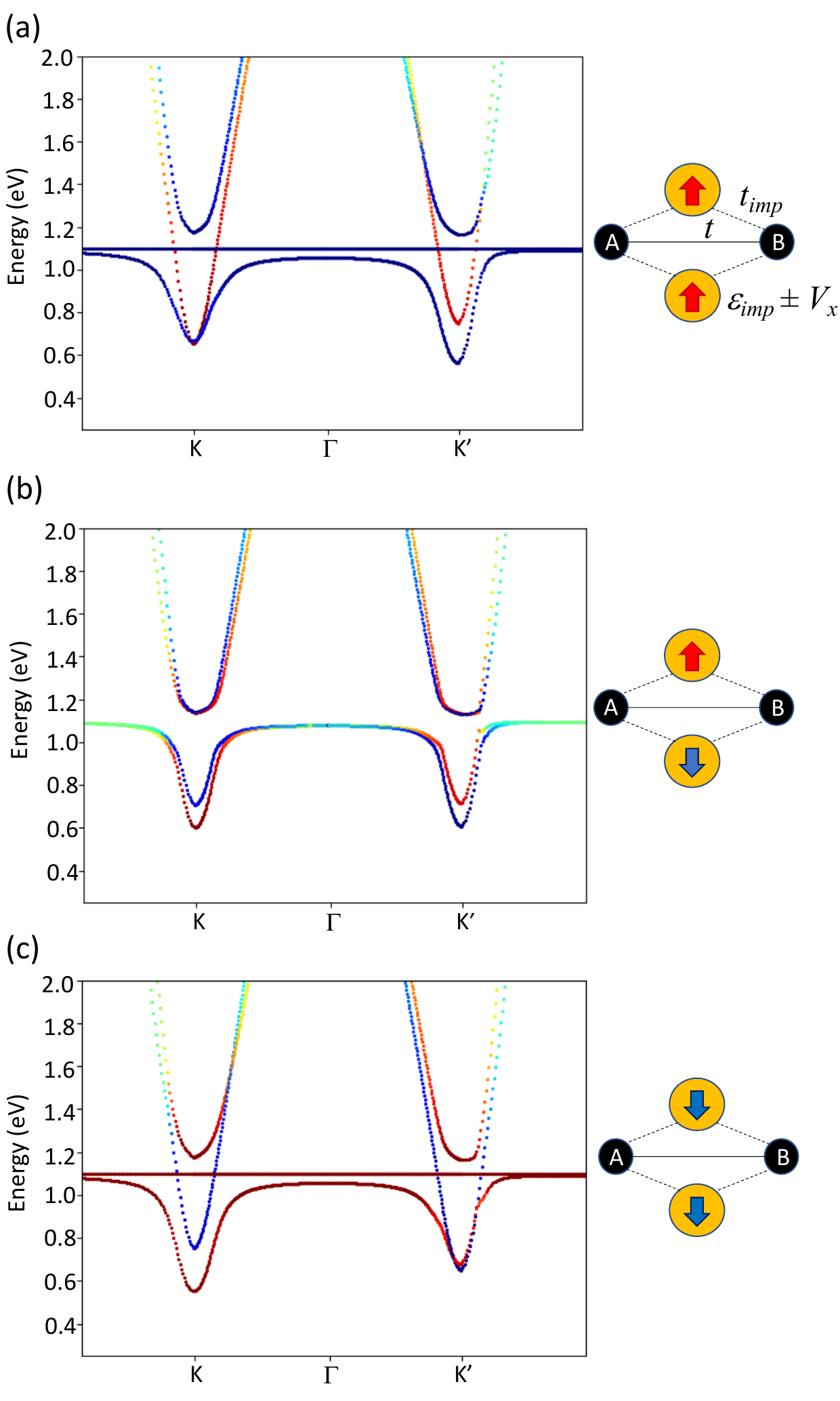}}
\caption{Description of the spin-valley symmetry breaking using the two impurity model Hamiltonian described in Eq.\ref{eqn:Ham}. When both impurities (magnetic layers) are spin-polarized in the up(down) direction, the TMD is polarized in the K'(K) valley (a,c). If both impurities (magnetic layers) are counter-polarized, the total magnetization in the TMD cancels and the spin-valley symmetry is restored (b).}
\label{fig_model}
\end{figure}

The results shown in Fig.\ref{fig_DFT} clearly highlight the importance of 2D magnetic materials in the design of future TMD-based van der Waals valleytronic devices. To better understand the physics underlying these results, and to give a prospect for the future design of van der Waals valleytronics, we use a toy model that contains the most important interactions giving rise to this big valley splitting. As a first approach to our valley-polarized device, we use a single-orbital tight-binding model in an hexagonal lattice with broken sublattice symmetry and spin-orbit coupling (SOC) to describe the electrons in the TMD. The encapsulating magnetic layers are described using two magnetic impurities which couple to the TMD with different magnetic configurations (see insets in Fig. \ref{fig_model}). Although CrBr$_3$ is a periodic two-dimensional material, the $t_{2g}$-bands of the Cr atoms involved in the valley splitting are almost dispersionless due to the small hopping between neighboring magnetic atoms. This justifies the use of magnetic impurities instead of a more realistic magnetic 2D material. The model Hamiltonian describing the van der Waals heterostructure is given by
\begin{eqnarray}
\label{eqn:Ham}
\mathcal{H} & = & \frac{1}{2}\sum_i\tau\epsilon_m c_i^\dagger c_i + t\sum_{\langle i,j \rangle} c_i^\dagger c_j \nonumber \\
& + & i\lambda_R\sum_{\langle i,j \rangle}c_i^\dagger(\sigma \times {\rm \mathbf{d_{ij}}})\cdot {\rm \mathbf{\hat{e}_z}}c_j \nonumber \\
& + & \frac{2i}{\sqrt{3}}\lambda_{SO}\sum_{\langle\langle i,j \rangle\rangle}c_i^\dagger\sigma\cdot\left({\rm \mathbf{d_{kj}}} \times {\rm \mathbf{d_{ik}}}\right)c_j \nonumber \\
& + & \sum_i (\epsilon_{imp} \pm \frac{V_x}{2}\sigma_z)d_i^\dagger d_i + t_{imp}\sum_{i,j}\left(c_i^\dagger d_j + d_i^\dagger c_j\right)
\end{eqnarray}
the first two terms describe electrons in a gapped hexagonal lattice, where $\tau=\pm 1$ is the sublattice index, $\epsilon_m$ is the mass term which opens up a gap in the band structure, $t$ is the hopping term between neighbouring $\langle i,j \rangle$ sites in the honeycomb lattice, and $c_i^\dagger,c_i$ are the creation and annihilation operators acting on each site of the hexagonal lattice. The third term introduces the spin-orbit Rashba coupling ($\lambda_R$) which is essential to couple spin and valley, with $\sigma$ the vector of Pauli matrices, $\rm \mathbf{d_{ij}}$ the unit vector connecting first neighbours in the honeycomb lattice, and $\rm \mathbf{\hat{e}_z}$ the unit cartesian vector along the $z$-direction. The fourth term is the intrinsic spin-orbit coupling ($\lambda_{SO}$), with $\rm \mathbf{d_{kj}}$ and $\rm \mathbf{d_{ik}}$ the two vectors connecting second-neighbor sites $\langle\langle i,j \rangle\rangle$. The fifth term includes the on-site energy ($\epsilon_{imp}$) and exchange interaction ($V_x$) for the magnetic layers where the $\pm$ sign controls the spin polarization, down or up, in the impurity. $\sigma_z$ is the $z$-component of the Pauli matrix vector, and $d_i^\dagger,d_i$ creates and annihilates an electron in the impurity sites. The last term allows to tune the hopping between the magnetic layers and the hexagonal lattice ($t_{imp}$).  

In Fig.\ref{fig_model}, we show the evolution of the valleys in the bottom of the conduction band of the TMD for different magnetic configurations of the magnetic layers. To model the FM/TMD/FM heterostructure we have used the Hamiltonian in Eq.\ref{eqn:Ham} with $\epsilon_m = 1.2$ eV, $t = 3$ eV, $\lambda_R = 0.09$ eV, $\lambda_{SO} = 0.3$ eV, $\epsilon_{imp} = 0.6$ eV, $V_x = 1$ eV, and $t_{imp} = 0.15$ eV (for more details, see Appendix \ref{AppC}). When the magnetic layers are co-polarized, an almost flat band from the CrBr$_3$ magnetic layers crosses the conduction bands of the TMD and hybridizes with the bands of the same spin, breaking the spin-valley symmetry and inducing a valley polarization that depends on the spin sign. Since the conduction bands in the model are mostly localized in the A-sublattice, $\epsilon_A = \epsilon_B+\epsilon_m$, the two impurities, i.e. magnetic layers, couple stronger to the A sites than to the B sites, leading to a weakly coupled flat band in the energy dispersion, as shown in Fig.\ref{fig_model}(a,c). For spin up(down), the TMD is K'(K)-polarized with a valley splitting of around $\sim 100$ meV as shown in Figs.\ref{fig_model}(a,c). When both magnetic layers are counter-polarized, spin up and down bands of the TMD are hybridized with the flat impurity bands, which eventually restores the spin-valley symmetry leading to a 0-polarized TMD as shown in Figs.\ref{fig_model} (b). Therefore, finding the correct combination of a TMD and a 2D magnetic semiconductor it is possible to efficiently break the spin-valley symmetry in the TMD leading to unexpectedly high values of valley splitting. 

\section{Multi-scale transport simulations.} 
\begin{figure}
\centerline{\includegraphics[width=\columnwidth]{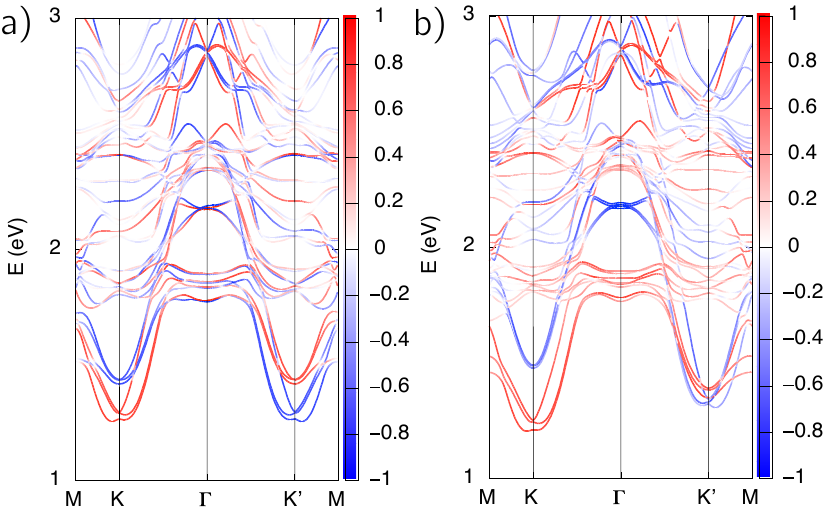}}
\caption{Band structure obtained from the wannierization of the  counter polarized (a) and co-polarized-down (b)  CrBr$_3$/WSe$_2$/CrBr$_3$ heterostructures}
\label{fig_SI_Wan}
\end{figure}

In order to precisely quantify and assess the potential performances of the proposed devices we perform precise multi-scale transport simulations. Upon the DFT simulation we perform the so-called wannierization to transform the Hamiltonian expressed in terms of plane-waves to maximally localized Wannier functions through the use of the Wannier90 code \cite{Wannier90}. 
The wannierization is performed projecting on the $\{\rm{d_{x^2-y^2},d_{z^2},d_{xy}}\}$ of the W atoms and on the $\{\rm{d_{xy},d_{xz},d_{yz}}\}$ of the Cr atoms of each layer, selecting as frozen window the energy range from 1.15 eV till 1.45 eV. In Fig.\ref{fig_SI_Wan} we report the spin-polarized band structure obtained after performing the wannierization for the counter polarized (a) and co-polarized-down (b) heterostructres. It's worth noting that this wannierization choice allows to retain an  excellent accuracy on the spin-polarized band structure while selecting only the energy window of interest for transport.

\begin{figure}[t]
\centerline{\includegraphics[width=\columnwidth]{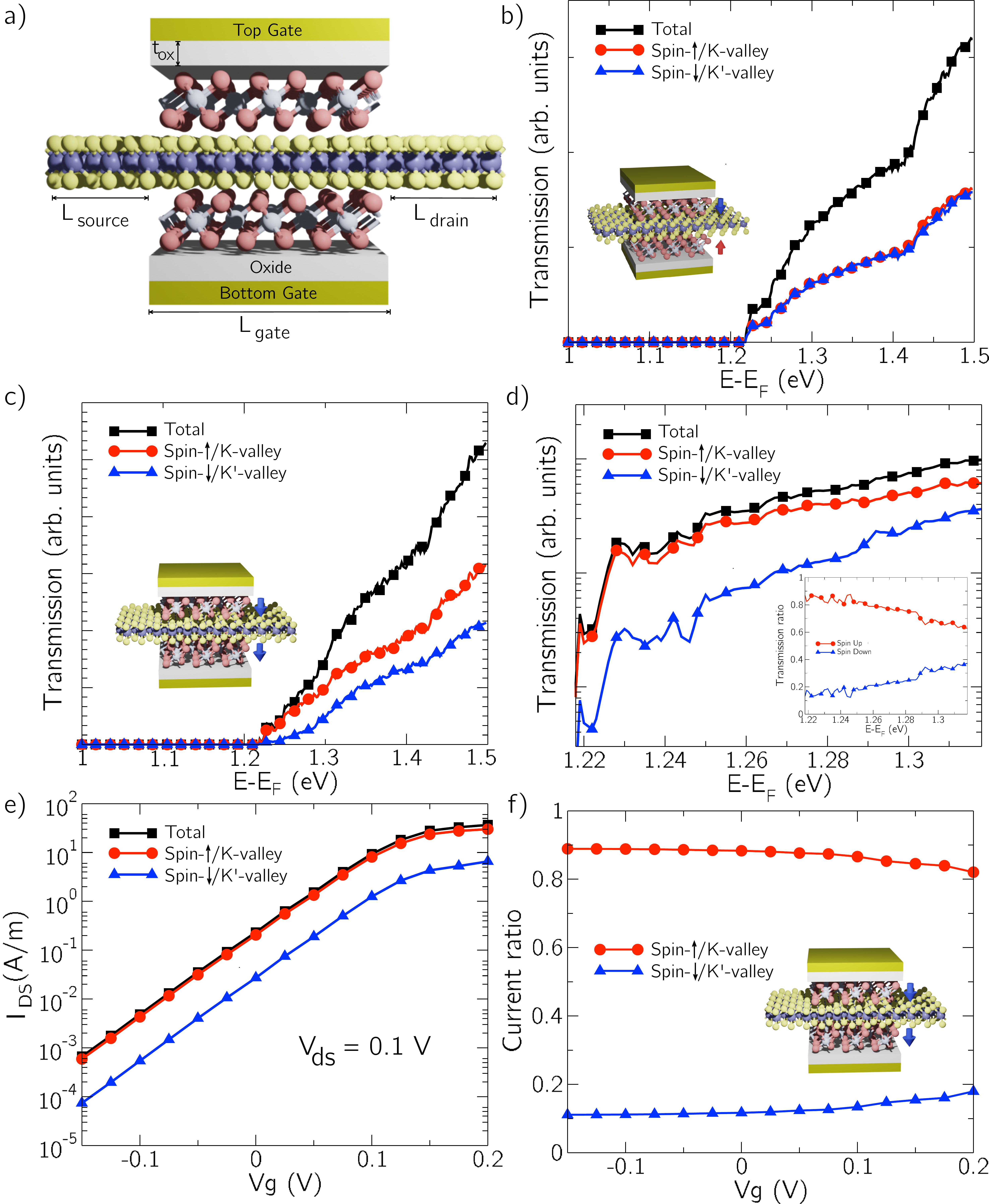}}
\caption{(a) Schematic depiction of the simulated device indicating the main geometrical parameters. Transmission coefficient for counter-polarized (b) and co-polarized down (c) channel. (d) Zoom of the transmission coefficient in semilogarithmic scale in the bottom of the conduction band (till 100 meV), with the transmission ratio in the same  range (inset). (e) Total and spin-valley polarized ($\uparrow$ and $\downarrow$ components) transfer characteristic for V$_{\text{ds}}$ = 0.1 V. (f) Current ratio extracted }
\label{fig_trans}
\end{figure}

The simulated device is schematically reported in Fig.\ref{fig_trans}(a), where we have considered as source and drain a region of monolayer WSe$_2$ while the channel is composed of CrBr$_3$/WSe$_2$/CrBr$_3$ heterostructure. Double (top and bottom) gate configuration is considered in the channel region, with an oxide between the gates and the channel material of 0.5 nm thickness and 3.9 relative dielectric constant. The channel length is 5.8 nm as well as the source and drain regions. In the source and drain contacts we have assume a doping of 5 x $10^{-3}$ molar fraction. In order to define the source and drain region with only WSe$_2$ monolayer, we have used the Wannier Hamiltonian obtained for the counter polarized structure in Fig.\ref{fig_SI_Wan}(a), following the procedure described in Ref.~\cite{Cannavo2021}, i.e. selecting from the heterostructure Hamiltonian only the sub-Hamiltonian belonging to the WSe$_2$ layer. We have ensured that the Hamiltonian of WSe$_2$ obtained from the procedure described in Ref.~\cite{Cannavo2021} is equal to the WSe$_2$ isolated monolayer. To ensure correct band allignment between source/drain and channel region, we have shift the onsite energies source and drain WSe$_2$ Hamiltonian of 70 meV. The energy shift has been calculated following the procedure described in Ref. \cite{Katagiri2016}.
In Figs.\ref{fig_trans}(b,c) we report the transmission coefficient around the minimum of the WSe$_2$ conduction band for flat potential while considering in the channel region counter-polarized, Fig.\ref{fig_trans}(b), and co-polarized-down, Fig.\ref{fig_trans}(c),  channel. We can observe that in the first case the total transmission is equally composed of both spin-$\uparrow$ (K-valley) and spin-$\downarrow$ (K'-valley) component, while in the case of co-polarized CrBr$_3$ layers we observe higher transmission for the spin-$\uparrow$ (K-valley) component. To better visualise it, we provide a zoom close to the bottom of the conduction band in Fig.\ref{fig_trans}(d). In the inset, we show the transmission ratio in the same energy range, where it is observed that the spin-$\uparrow$/K-valley contributes 80$\%$-60$\%$ of the total transmission. Finally, we perform fully transport simulations of the device solving self-consistently non-equilibrium Green's functions and the Poisson equation using NanoTCAD ViDES sofware \cite{vides,Vides2023}. In Fig.\ref{fig_trans}(e) we report the transfer characteristic of the device while applying gate voltages from -0.15 V till 0.2 V and fixing a source-to-drain voltage of 0.1 V. We can observe that the device switch OFF/ON while maintaining a predominance of spin-$\uparrow$/K-valley current component. To quantify the degree of spin/valley current polarization we have computed the current ratio in Fig.\ref{fig_trans}(f), where we can observe that in the operation range considered the current spin/valley polarization is always higher than 80$\%$.

\section{Conclusions} 
In this work, we have shown by means of DFT and transport simulations the possibility to engineer valleytronic devices out of van der Waals magnetic heterostructures combining TMDs and 2D magnetic materials. In the particular case studied, namely CrBr$_3$/WSe$_2$/CrBr$_3$ trilayer, we find a giant valley splitting of the order of 100 meV in the conduction band of the TMD under compressive strain due to the hybridization of the $t_{2g}$ bands of CrBr$_3$ with the empty $d_{xy}$ and $d_{x^2-y^2}$ bands of WSe$_2$. The transport simulations based on DFT show a valley polarization of the injected current that remains for gate voltage range of 0.3 V. This work lays the foundations for the future design of valleytronic devices based on van der Waals heterostructures.  

\section*{Acknowledgments} 
D.S. acknowledges financial support from Generalitat Valenciana through the CIDEGENT program (Nr. CIDEGENT/2021/052). This study forms part of the Advanced Materials program and was supported by MCIN with funding from European Union NextGenerationEU (PRTR-C17.I1) and by Generalitat Valenciana (MFA/2022/045). The authors gratefully acknowledge Italian Ministry of University and Research (MUR) PRIN project SECSY (2022FPAKWF) and CN00000013 National Centre for HPC, Big Data and Quantum Computing (HPC). 

\appendix

\section{Spin-valley splitting in Bilayer-C\lowercase{r}B\lowercase{r}$_3$/WS\lowercase{e}$_2$/C\lowercase{r}B\lowercase{r}$_3$ heterostructure and effect of perpendicular electric field}
\label{AppA}

The proof-of-principle device reported in Fig.\ref{fig_device} is made of WSe$_2$-layer encapsulated on top by a monolayer CrBr$_3$ and on bottom by a bilayer of CrBr$_3$ in the monoclinic stacking and relies on the possibility of switching the interlayer magnetic order of layer I and II by the top and bootom gate. Nevertheless, in the subsequent calculations we have focused our attention on the top three layers, where the proximity effects play a key role, and we have assumed that the electric field produced by the gates do not alter the valley splitting. In the present Appendix we justify both assumptions. 

\begin{figure}[h!!!]
\centerline{\includegraphics[width=0.9\columnwidth]{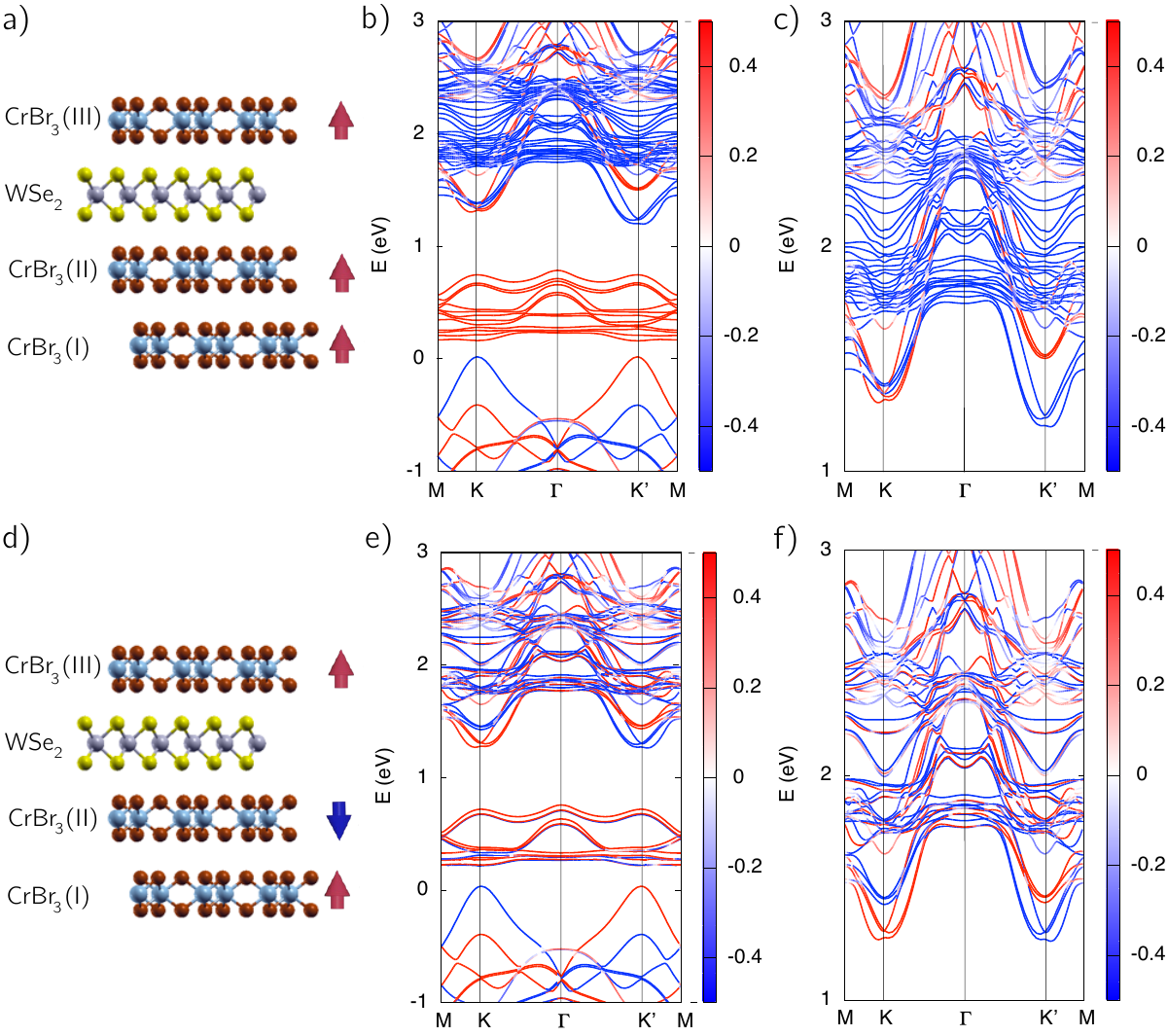}}
\caption{(a) Lateral view of the crystalline structure of the four layers structure composed of a bottom CrBr$_3$ bilayer, WSe$_2$ and monolayer CrBr$_3$ all the magnetic layers are co-polarized. (b) Band structure of the co-polarized Bilayer-CrBr$_3$/WSe$_2$/CrBr$_3$ heterostructure and (c) relative zoom near the bottom of the WSe$_2$ conduction band. (d) Lateral view of the crystalline structure of the four layers structure composed of a bottom CrBr$_3$ bilayer, WSe$_2$ and monolayer CrBr$_3$, the layers CrBr$_3$ (II) and CrBr$_3$ (III) are counter-polarized. (e) Band structure of the counter-polarized Bilayer-CrBr$_3$/WSe$_2$/CrBr$_3$ heterostructure and (f) relative zoom near the bottom of the WSe$_2$ conduction band. }
\label{fig_SI_4layers}
\end{figure}

Firstly, we have fully relaxed the bilayer-C\lowercase{r}B\lowercase{r}$_3$/WS\lowercase{e}$_2$/C\lowercase{r}B\lowercase{r}$_3$ heterostructre, with the same parameters reported in the main text. We have observed that the CrBr$_3$ (II) layer is slightly closer to the WSe$_2$ layer (0.007 nm) respect to the top layer. To demonstrate the negligible influence of the first bottom layer (CrBr$_3$ (I)) on the valley-splitting of the WSe$_2$ conduction band, in Fig.\ref{fig_SI_4layers}(b) we report the spin-projected band structure and a zoom (Fig.\ref{fig_SI_4layers}(c)) near the bottom of the conduction band of the WSe$_2$ for the co-polarized structure reported in Fig.\ref{fig_SI_4layers}(a). We observe that the valley splitting is preserved, if not slightly enhanced (we measure a splitting of 107 meV): the enhancement is mainly due to the fact that the CrBr$_3$ (II) is closer to the WSe$_2$ layer. In Fig.\ref{fig_SI_4layers}(e) we report the band structure and the zoom (Fig.\ref{fig_SI_4layers}(f)) near the bottom of the conduction band of the four-layers structure but with the CrBr$_3$ layers (II) and (III) counter-polarized. In this case, no valley-splitting is observed, so the bottom layer CrBr$_3$ (I), also for this configuration, does not alter the bottom of the conduction band of the WSe$_2$.

\begin{figure}[h!!!]
\centerline{\includegraphics[width=0.6\columnwidth]{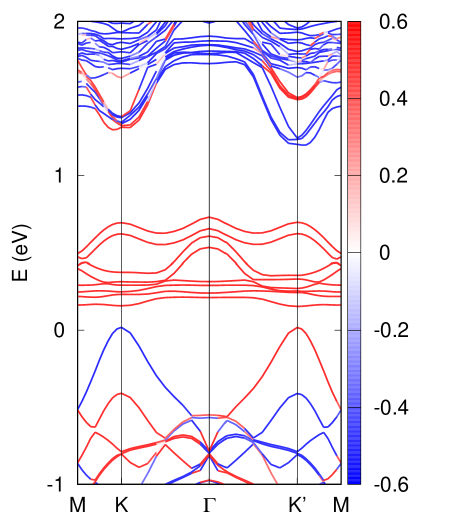}}
\caption{Band structure of CrBr$_3$/WSe$_2$/CrBr$_3$ heterostructure with an external electric field $\mathcal{E}$ = 0.8 V/nm applied perpendicular to the layers.}
\label{fig_SI_Field}
\end{figure}

Once we have verify that the bottom CrBr$_3$ (I) does not influence the valley splitting, we have carried out a DFT calculation on the trilayer heterostructure in order to discard detrimental effects on the valley splitting due to the electric field produced by the gates. In particular we have considered a perpendicular electric field $\mathcal{E}$= 0.8 V/nm. The results are shown in Fig.\ref{fig_SI_Field}.

In Fig.\ref{fig_SI_Field} we clearly see how the electric field splits the $e_g$ bands of the CrBr$_3$ over the Fermi energy. However, the valley splitting in the conduction band is not affected by the electric field and remains almost of the same value (99 meV). Since the empty $t_{2g}$ bands of the CrBr$_3$ layers crossing the conduction band of WSe$_2$ are not strongly affected by the electric field, the hybridization between $d$-bands is maintained, leading to the same valley splitting even in presence of the electric field.

\section{Spin-valley splitting in other heterostructures}
\label{AppB}

\begin{figure}[t]
\centerline{\includegraphics[width=\columnwidth]{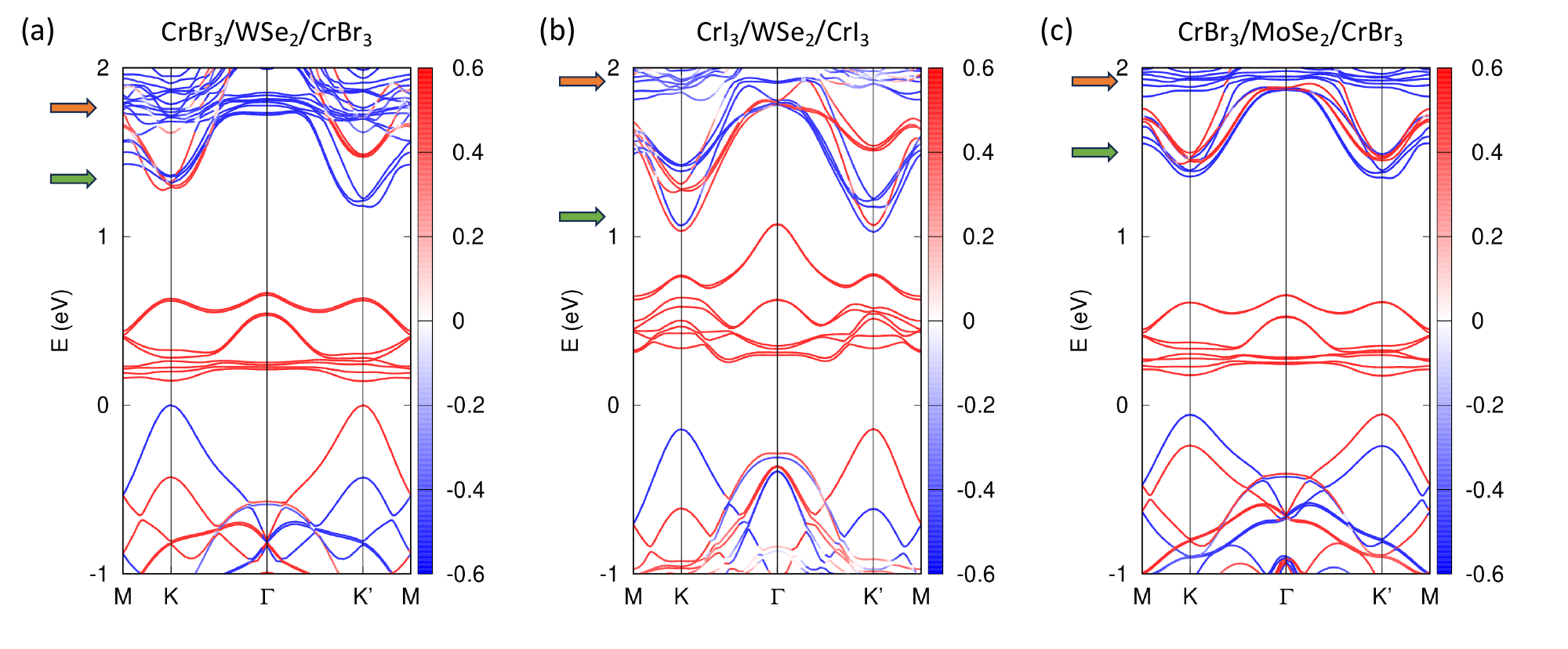}}
\caption{Band structure of (a) CrBr$_3$/WSe$_2$/CrBr$_3$, (b) CrI$_3$/WSe$_2$/CrI$_3$, and (c) 
 CrBr$_3$/MoSe$_2$/CrBr$_3$ heterostructures. Orange and green arrows indicate the position of $t_{2g}$ bands and the bottom of the conduction bands of the TMD respectively. }
\label{fig_SI_hetero}
\end{figure}

To highlight the importance of the band alignment and spin-orbit coupling on the spin-valley splitting mechanism in magnetic van der Waals heterostructures, in Fig.\ref{fig_SI_hetero} we have compared electronic structure of the CrBr$_3$/WSe$_2$/CrBr$_3$ heterostructure with other two heterostructures, namely CrI$_3$/WSe$_2$/CrI$_3$ and CrBr$_3$/MoSe$_2$/CrBr$_3$. In this figure, it is possible to compare the position of the $t_{2g}$ bands of the magnetic layers (orange arrow) with respect to the bottom of the conduction band of the TMD (green arrow). The spin-valley splitting in the conduction band is only observed for the case of CrBr$_3$/WSe$_2$/CrBr$_3$ due to the proximity of the $t_{2g}$ bands of the CrBr$_3$ to the bottom of the conduction band of WSe$_2$. In the case of CrBr$_3$/MoSe$_2$/CrBr$_3$, although the alignment is good, the lower spin-orbit coupling (clearly denoted by the smaller splitting in the valence bands compared with WSe$_2$) reduces the valley splitting. It is interesting to observe in this case that in both K and K' the conduction bands are polarized down (blue color) in contrast to the other two cases.   

\section{Single-orbital tight-binding model of a magnetically-encapsulated transition metal dichalcogenide}
\label{AppC}

Several parameters are involved in the description of the electronic structure of a magnetically-encapsulated transition metal dichalcogenide (TMD). In this SI, we focus first on the effect of the Rashba splitting ($\lambda_R$) and the intrinsic spin-orbit coupling ($\lambda_{SO}$) to properly tune the spin-splitting and the expectation value of the off-plane spin operator $\langle S_z \rangle$ of the TMD conduction bands. Then, we show how the intensity of the exchange interaction of the ferromagnets (FM) affects the electronic structure of the TMD for a constant FM-TMD interaction. Finally, we study the tunability of the valley splitting for different values of the FM-TMD interaction. 

\begin{figure}[h]
\centerline{\includegraphics[width=\columnwidth]{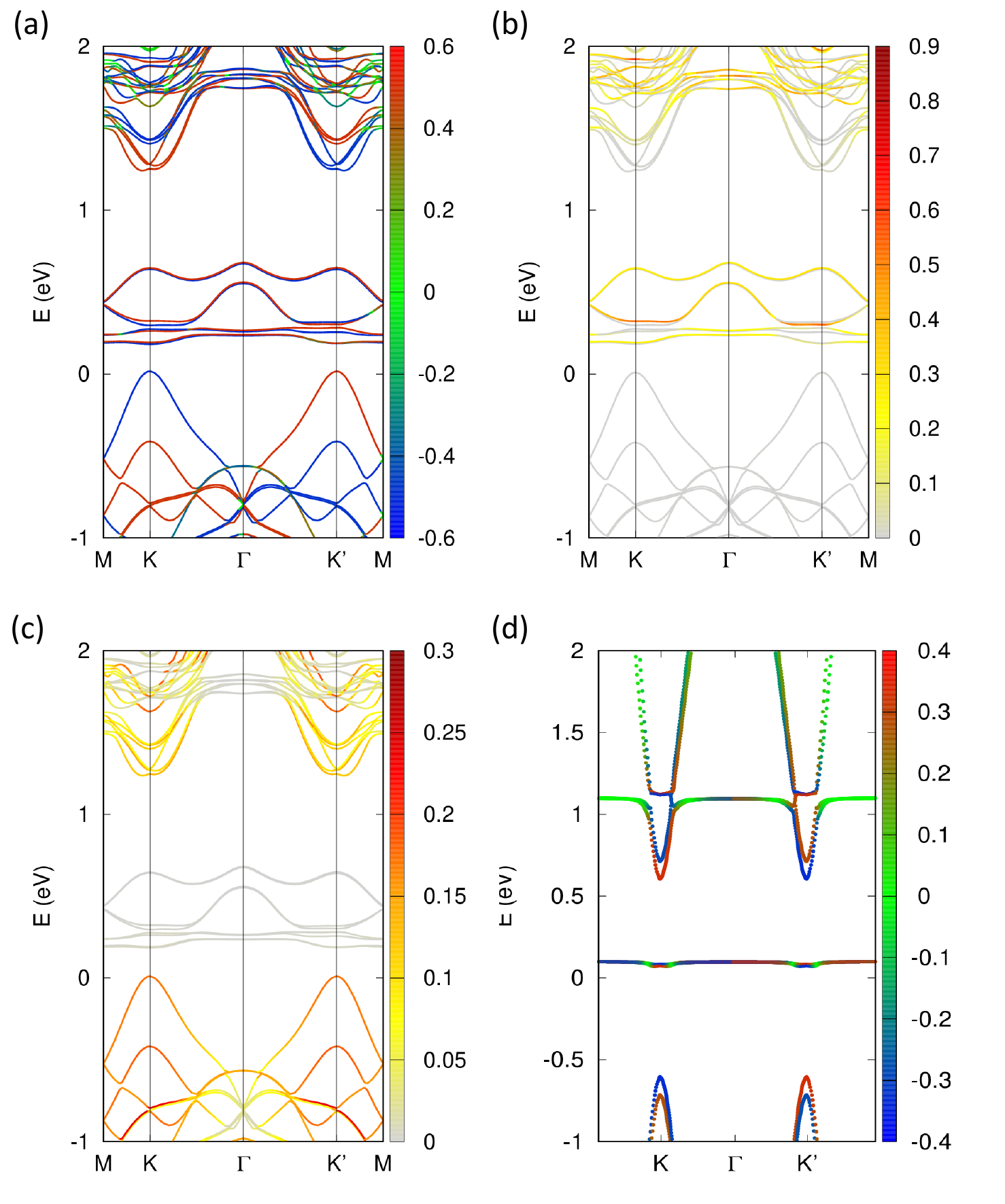}}
\caption{(a) Spin-projected band structure of the CrBr$_3$/WSe$_2$/CrBr$_3$ heterostructure with antiparallel magnetic layers. The bar color represents the expectation value of the $S_z$ operator. (b) The band structure projected on $d$-orbitals of a single Cr atom in one magnetic layer. (c) Idem for the W $d$-orbitals. (d) Bands obtained with the effective model. The color bar in (a) and (d) represents the expectation value of the $z$-component of the spin $\langle S_z \rangle$. In (b) and (c) represent the spectral weight on the $d$-orbitals in Cr and W respectively.}
\label{fig_SI_AFM}
\end{figure}

{\bf Spin-orbit coupling ($\lambda_r$ and $\lambda_{SO}$).} In Fig.\ref{fig_SI_AFM}(a), we show the spin-projected bands of the CrBr$_3$/WSe$_2$/CrBr$_3$ heterostructure with antiferromagnetically(AFM)-aligned magnetic layers. This is the starting point for our tight-binding model (Fig.\ref{fig_SI_AFM}(d)). It shows the typical gapped TMD band structure with the two valleys at K and K', the high spin splitting in the valence bands, and the small spin-splitting in the conduction bands (spin up and down bands are plotted in red and blue respectively). On top of the TMD band structure, it is possible to see several flat bands crossing at around $0.5$ eV above the bottom of the conduction bands of WSe$_2$ belonging to the CrBr$_3$ layers. These bands can be easily observed in the band projections shown in Fig.\ref{fig_SI_AFM}(b,c) around $E = 1.8$ eV at the $\Gamma$-point.  

\begin{figure}[t!]
\centerline{\includegraphics[width=\columnwidth]{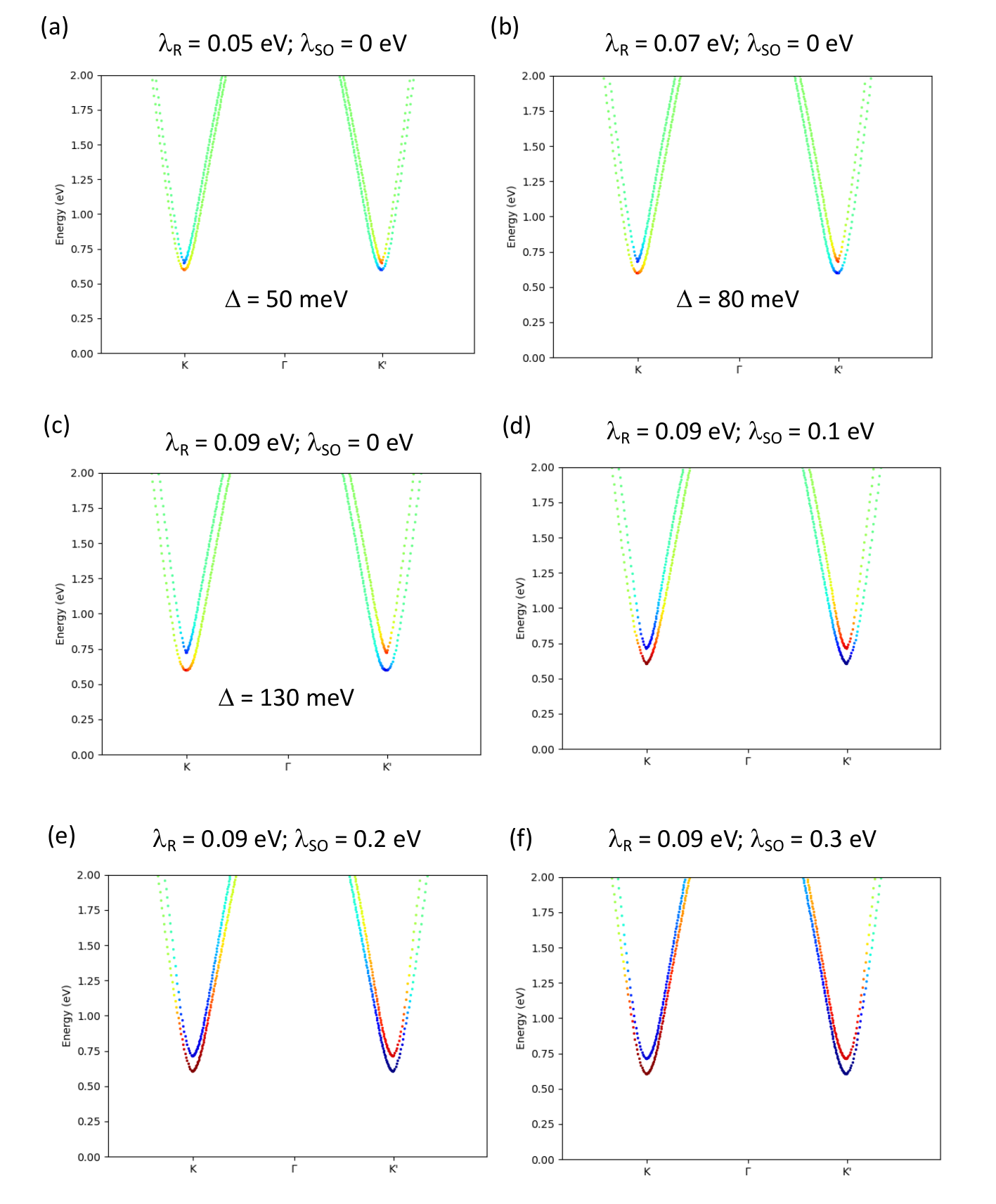}}
\caption{(a-c) Evolution of the TMD conduction band splitting with increasing values of the Rashba parameter. (d-f) Effect of the intrinsic spin-orbit coupling in the $S_z$ expectation value along the conduction bands.}
\label{fig_SI_SOC}
\end{figure}

The measured spin-splitting in the conduction bands of the WSe$_2$ with AFM-aligned magnetic layers is of the order of $\Delta = 125$ meV, much higher than previously reported values \cite{KosmiderRossier2013}. In Fig.\ref{fig_SI_SOC}(a-c), we show the evolution of the conduction band of the TMD for different values of the Rashba splitting using the tight-binding model in Eq. 1 of the main manuscript. For $\lambda_R = 90$ meV, we find a splitting of $\Delta = 130$ meV, very similar to the DFT result. The intrinsic spin-orbit coupling has important implications in the spin texture around the valleys and competes with the Rashba SOC \cite{CummingsRoche2017}. While the Rashba coupling contributes to a more planar and helical spin texture around the valleys, the intrinsic spin-orbit coupling tends to align the spins along the off-plane direction. In Fig.\ref{fig_SI_SOC}(d-f), we show how the in-plane contribution (green dots) is reduced along the K-$\Gamma$-K' path for increasing values of the intrinsic spin-orbit coupling, and the bands become up (red) and down (blue) spin-polarized. As shown in Fig.\ref{fig_SI_AFM}(a), the valence and conduction bands of the TMD are mostly off-plane polarized with a very small in-plane contribution. In our model, we find this situation for $\lambda_{SO} = 300$ meV, which is close to the reported atomic spin-orbit coupling of the Se atom $\lambda_{SO} = 220$ meV \cite{WittelManne1974}.             

\begin{figure}[t]
\centerline{\includegraphics[width=\columnwidth]{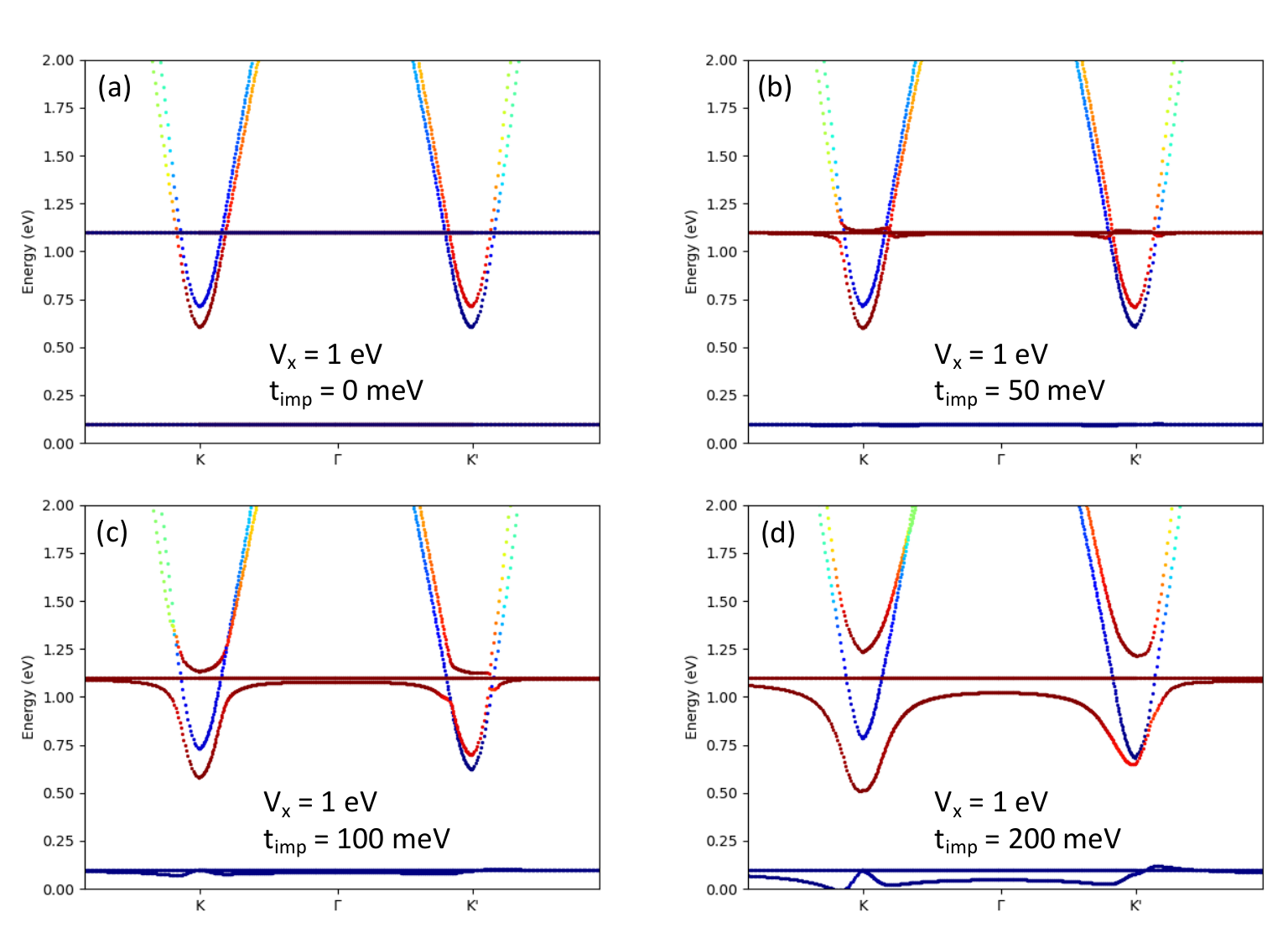}}
\caption{(a) Model conduction bands with antiferromagnetically aligned  impurities using $\epsilon_{imp} = 0.6$ and $V_x = 1$ eV. Model conduction bands with ferromagnetically aligned magnetic impurities for (b) $t_{imp} = 50$, (c) $t_{imp} = 100$, and (d) $t_{imp} = 200$ meV.}
\label{fig_SI_exchange}
\end{figure}

{\bf Exchange interaction ($V_{x}$).} Exchange interaction is a key ingredient, together with SOC, to realize a valley splitting. In our heterostructure, the exchange field comes from the CrBr$_3$ encapsulated layers. When the magnetization is turned on, the $e_g$ bands split $\sim 1$ eV, as shown in Fig.\ref{fig_SI_exchange}(a) for the AFM case. In our model, we have located the impurity band at the bottom of the conduction band $\epsilon_{imp} = 0.6$ eV with an exchange interaction $V_x = 1$ eV (see Fig.\ref{fig_SI_exchange}(b)). 

\begin{figure}
\centerline{\includegraphics[width=\columnwidth]{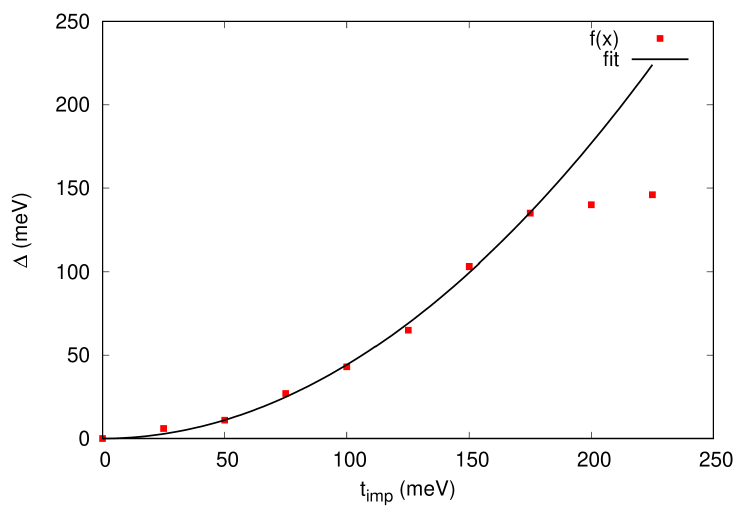}}
\caption{Evolution of the valley splitting with increasing interlayer interaction (red squares). The fitting to a parabolic dispersion is shown in black.}
\label{fig_SI_fit}
\end{figure}

{\bf FM-TMD interaction ($t_{imp}$).} Finally, we focus on the interlayer coupling between the ferromagnetic encapsulating layers and the TMD. The interaction between the orbitals located at different layers is crucial for the observation of a valley polarization. In Fig.\ref{fig_SI_exchange}(b-d), we show the evolution of valley splitting for three different interlayer hopping parameter, namely, $t_{imp} = 50$, $100$, and $200$ meV. To have a more clear picture of the evolution of the valley splitting with the interlayer hopping, we have calculated the splitting for several values of the interlayer hopping $t_{imp}$ (see Fig.\ref{fig_SI_fit}). For $t_{imp} \leq 200$ meV the evolution is parabolic $\Delta(t_{imp}) = 0.004t_{imp}^2$, and then saturates for $t_{imp} > 200$ meV.   

\begin{thebibliography}{39}%
\makeatletter
\providecommand \@ifxundefined [1]{%
 \@ifx{#1\undefined}
}%
\providecommand \@ifnum [1]{%
 \ifnum #1\expandafter \@firstoftwo
 \else \expandafter \@secondoftwo
 \fi
}%
\providecommand \@ifx [1]{%
 \ifx #1\expandafter \@firstoftwo
 \else \expandafter \@secondoftwo
 \fi
}%
\providecommand \natexlab [1]{#1}%
\providecommand \enquote  [1]{``#1''}%
\providecommand \bibnamefont  [1]{#1}%
\providecommand \bibfnamefont [1]{#1}%
\providecommand \citenamefont [1]{#1}%
\providecommand \href@noop [0]{\@secondoftwo}%
\providecommand \href [0]{\begingroup \@sanitize@url \@href}%
\providecommand \@href[1]{\@@startlink{#1}\@@href}%
\providecommand \@@href[1]{\endgroup#1\@@endlink}%
\providecommand \@sanitize@url [0]{\catcode `\\12\catcode `\$12\catcode
  `\&12\catcode `\#12\catcode `\^12\catcode `\_12\catcode `\%12\relax}%
\providecommand \@@startlink[1]{}%
\providecommand \@@endlink[0]{}%
\providecommand \url  [0]{\begingroup\@sanitize@url \@url }%
\providecommand \@url [1]{\endgroup\@href {#1}{\urlprefix }}%
\providecommand \urlprefix  [0]{URL }%
\providecommand \Eprint [0]{\href }%
\providecommand \doibase [0]{https://doi.org/}%
\providecommand \selectlanguage [0]{\@gobble}%
\providecommand \bibinfo  [0]{\@secondoftwo}%
\providecommand \bibfield  [0]{\@secondoftwo}%
\providecommand \translation [1]{[#1]}%
\providecommand \BibitemOpen [0]{}%
\providecommand \bibitemStop [0]{}%
\providecommand \bibitemNoStop [0]{.\EOS\space}%
\providecommand \EOS [0]{\spacefactor3000\relax}%
\providecommand \BibitemShut  [1]{\csname bibitem#1\endcsname}%
\let\auto@bib@innerbib\@empty
\bibitem [{\citenamefont {Cao}\ \emph {et~al.}(2018)\citenamefont {Cao},
  \citenamefont {Fatemi}, \citenamefont {Fang}, \citenamefont {Watanabe},
  \citenamefont {Taniguchi}, \citenamefont {Kaxiras},\ and\ \citenamefont
  {Jarillo-Herrero}}]{CaoJarillo2018}%
  \BibitemOpen
  \bibfield  {author} {\bibinfo {author} {\bibfnamefont {Y.}~\bibnamefont
  {Cao}}, \bibinfo {author} {\bibfnamefont {V.}~\bibnamefont {Fatemi}},
  \bibinfo {author} {\bibfnamefont {S.}~\bibnamefont {Fang}}, \bibinfo {author}
  {\bibfnamefont {K.}~\bibnamefont {Watanabe}}, \bibinfo {author}
  {\bibfnamefont {T.}~\bibnamefont {Taniguchi}}, \bibinfo {author}
  {\bibfnamefont {E.}~\bibnamefont {Kaxiras}},\ and\ \bibinfo {author}
  {\bibfnamefont {P.}~\bibnamefont {Jarillo-Herrero}},\ }\bibfield  {title}
  {\bibinfo {title} {Unconventional superconductivity in magic-angle graphene
  superlattices},\ }\href {https://doi.org/10.1038/nature26160} {\bibfield
  {journal} {\bibinfo  {journal} {Nature}\ }\textbf {\bibinfo {volume} {556}},\
  \bibinfo {pages} {43} (\bibinfo {year} {2018})}\BibitemShut {NoStop}%
\bibitem [{\citenamefont {Song}\ \emph {et~al.}(2018)\citenamefont {Song},
  \citenamefont {Soriano}, \citenamefont {Cummings}, \citenamefont {Robles},
  \citenamefont {Ordejón},\ and\ \citenamefont {Roche}}]{SongRoche2018}%
  \BibitemOpen
  \bibfield  {author} {\bibinfo {author} {\bibfnamefont {K.}~\bibnamefont
  {Song}}, \bibinfo {author} {\bibfnamefont {D.}~\bibnamefont {Soriano}},
  \bibinfo {author} {\bibfnamefont {A.~W.}\ \bibnamefont {Cummings}}, \bibinfo
  {author} {\bibfnamefont {R.}~\bibnamefont {Robles}}, \bibinfo {author}
  {\bibfnamefont {P.}~\bibnamefont {Ordejón}},\ and\ \bibinfo {author}
  {\bibfnamefont {S.}~\bibnamefont {Roche}},\ }\bibfield  {title} {\bibinfo
  {title} {Spin proximity effects in graphene/topological insulator
  heterostructures},\ }\href {https://doi.org/10.1021/acs.nanolett.7b05482}
  {\bibfield  {journal} {\bibinfo  {journal} {Nano Letters}\ }\textbf {\bibinfo
  {volume} {18}},\ \bibinfo {pages} {2033} (\bibinfo {year}
  {2018})}\BibitemShut {NoStop}%
\bibitem [{\citenamefont {Karpiak}\ \emph {et~al.}(2019)\citenamefont
  {Karpiak}, \citenamefont {Cummings}, \citenamefont {Zollner}, \citenamefont
  {Vila}, \citenamefont {Khokhriakov}, \citenamefont {Hoque}, \citenamefont
  {Dankert}, \citenamefont {Svedlindh}, \citenamefont {Fabian}, \citenamefont
  {Roche},\ and\ \citenamefont {Dash}}]{KarpiakDash2019}%
  \BibitemOpen
  \bibfield  {author} {\bibinfo {author} {\bibfnamefont {B.}~\bibnamefont
  {Karpiak}}, \bibinfo {author} {\bibfnamefont {A.~W.}\ \bibnamefont
  {Cummings}}, \bibinfo {author} {\bibfnamefont {K.}~\bibnamefont {Zollner}},
  \bibinfo {author} {\bibfnamefont {M.}~\bibnamefont {Vila}}, \bibinfo {author}
  {\bibfnamefont {D.}~\bibnamefont {Khokhriakov}}, \bibinfo {author}
  {\bibfnamefont {A.~M.}\ \bibnamefont {Hoque}}, \bibinfo {author}
  {\bibfnamefont {A.}~\bibnamefont {Dankert}}, \bibinfo {author} {\bibfnamefont
  {P.}~\bibnamefont {Svedlindh}}, \bibinfo {author} {\bibfnamefont
  {J.}~\bibnamefont {Fabian}}, \bibinfo {author} {\bibfnamefont
  {S.}~\bibnamefont {Roche}},\ and\ \bibinfo {author} {\bibfnamefont {S.~P.}\
  \bibnamefont {Dash}},\ }\bibfield  {title} {\bibinfo {title} {Magnetic
  proximity in a van der waals heterostructure of magnetic insulator and
  graphene},\ }\href {https://doi.org/10.1088/2053-1583/ab5915} {\bibfield
  {journal} {\bibinfo  {journal} {2D Materials}\ }\textbf {\bibinfo {volume}
  {7}},\ \bibinfo {pages} {015026} (\bibinfo {year} {2019})}\BibitemShut
  {NoStop}%
\bibitem [{\citenamefont {Zhong}\ \emph {et~al.}(2020)\citenamefont {Zhong},
  \citenamefont {Seyler}, \citenamefont {Linpeng}, \citenamefont {Wilson},
  \citenamefont {Taniguchi}, \citenamefont {Watanabe}, \citenamefont {McGuire},
  \citenamefont {Fu}, \citenamefont {Xiao}, \citenamefont {Yao},\ and\
  \citenamefont {Xu}}]{ZhongXu2020}%
  \BibitemOpen
  \bibfield  {author} {\bibinfo {author} {\bibfnamefont {D.}~\bibnamefont
  {Zhong}}, \bibinfo {author} {\bibfnamefont {K.~L.}\ \bibnamefont {Seyler}},
  \bibinfo {author} {\bibfnamefont {X.}~\bibnamefont {Linpeng}}, \bibinfo
  {author} {\bibfnamefont {N.~P.}\ \bibnamefont {Wilson}}, \bibinfo {author}
  {\bibfnamefont {T.}~\bibnamefont {Taniguchi}}, \bibinfo {author}
  {\bibfnamefont {K.}~\bibnamefont {Watanabe}}, \bibinfo {author}
  {\bibfnamefont {M.~A.}\ \bibnamefont {McGuire}}, \bibinfo {author}
  {\bibfnamefont {K.-M.~C.}\ \bibnamefont {Fu}}, \bibinfo {author}
  {\bibfnamefont {D.}~\bibnamefont {Xiao}}, \bibinfo {author} {\bibfnamefont
  {W.}~\bibnamefont {Yao}},\ and\ \bibinfo {author} {\bibfnamefont
  {X.}~\bibnamefont {Xu}},\ }\bibfield  {title} {\bibinfo {title}
  {Layer-resolved magnetic proximity effect in van der waals
  heterostructures},\ }\href {https://doi.org/10.1038/s41565-019-0629-1}
  {\bibfield  {journal} {\bibinfo  {journal} {Nature Nanotechnology}\ }\textbf
  {\bibinfo {volume} {15}},\ \bibinfo {pages} {187} (\bibinfo {year}
  {2020})}\BibitemShut {NoStop}%
\bibitem [{\citenamefont {{Kezilebieke}}\ \emph {et~al.}(2020)\citenamefont
  {{Kezilebieke}}, \citenamefont {{Nurul Huda}}, \citenamefont {{Va{\v{n}}o}},
  \citenamefont {{Aapro}}, \citenamefont {{Ganguli}}, \citenamefont
  {{Silveira}}, \citenamefont {{G{\l}odzik}}, \citenamefont {{Foster}},
  \citenamefont {{Ojanen}},\ and\ \citenamefont
  {{Liljeroth}}}]{KezilebiekeLiljeroth2020}%
  \BibitemOpen
  \bibfield  {author} {\bibinfo {author} {\bibfnamefont {S.}~\bibnamefont
  {{Kezilebieke}}}, \bibinfo {author} {\bibfnamefont {M.}~\bibnamefont {{Nurul
  Huda}}}, \bibinfo {author} {\bibfnamefont {V.}~\bibnamefont {{Va{\v{n}}o}}},
  \bibinfo {author} {\bibfnamefont {M.}~\bibnamefont {{Aapro}}}, \bibinfo
  {author} {\bibfnamefont {S.~C.}\ \bibnamefont {{Ganguli}}}, \bibinfo {author}
  {\bibfnamefont {O.~J.}\ \bibnamefont {{Silveira}}}, \bibinfo {author}
  {\bibfnamefont {S.}~\bibnamefont {{G{\l}odzik}}}, \bibinfo {author}
  {\bibfnamefont {A.~S.}\ \bibnamefont {{Foster}}}, \bibinfo {author}
  {\bibfnamefont {T.}~\bibnamefont {{Ojanen}}},\ and\ \bibinfo {author}
  {\bibfnamefont {P.}~\bibnamefont {{Liljeroth}}},\ }\bibfield  {title}
  {\bibinfo {title} {{Topological superconductivity in a van der Waals
  heterostructure}},\ }\href {https://doi.org/10.1038/s41586-020-2989-y}
  {\bibfield  {journal} {\bibinfo  {journal} {Nature}\ }\textbf {\bibinfo
  {volume} {588}},\ \bibinfo {pages} {424} (\bibinfo {year}
  {2020})}\BibitemShut {NoStop}%
\bibitem [{\citenamefont {Lyons}\ \emph {et~al.}(2020)\citenamefont {Lyons},
  \citenamefont {Gillard}, \citenamefont {Molina-S{\'a}nchez}, \citenamefont
  {Misra}, \citenamefont {Withers}, \citenamefont {Keatley}, \citenamefont
  {Kozikov}, \citenamefont {Taniguchi}, \citenamefont {Watanabe}, \citenamefont
  {Novoselov}, \citenamefont {Fern{\'a}ndez-Rossier},\ and\ \citenamefont
  {Tartakovskii}}]{LyonsTartakovskii2020}%
  \BibitemOpen
  \bibfield  {author} {\bibinfo {author} {\bibfnamefont {T.~P.}\ \bibnamefont
  {Lyons}}, \bibinfo {author} {\bibfnamefont {D.}~\bibnamefont {Gillard}},
  \bibinfo {author} {\bibfnamefont {A.}~\bibnamefont {Molina-S{\'a}nchez}},
  \bibinfo {author} {\bibfnamefont {A.}~\bibnamefont {Misra}}, \bibinfo
  {author} {\bibfnamefont {F.}~\bibnamefont {Withers}}, \bibinfo {author}
  {\bibfnamefont {P.~S.}\ \bibnamefont {Keatley}}, \bibinfo {author}
  {\bibfnamefont {A.}~\bibnamefont {Kozikov}}, \bibinfo {author} {\bibfnamefont
  {T.}~\bibnamefont {Taniguchi}}, \bibinfo {author} {\bibfnamefont
  {K.}~\bibnamefont {Watanabe}}, \bibinfo {author} {\bibfnamefont {K.~S.}\
  \bibnamefont {Novoselov}}, \bibinfo {author} {\bibfnamefont {J.}~\bibnamefont
  {Fern{\'a}ndez-Rossier}},\ and\ \bibinfo {author} {\bibfnamefont {A.~I.}\
  \bibnamefont {Tartakovskii}},\ }\bibfield  {title} {\bibinfo {title}
  {{Interplay between spin proximity effect and charge-dependent exciton
  dynamics in MoSe$_2$/CrBr$_3$ van der Waals heterostructures}},\ }\href
  {https://doi.org/10.1038/s41467-020-19816-4} {\bibfield  {journal} {\bibinfo
  {journal} {Nat. Commun.}\ }\textbf {\bibinfo {volume} {11}},\ \bibinfo
  {pages} {6021} (\bibinfo {year} {2020})}\BibitemShut {NoStop}%
\bibitem [{\citenamefont {Gong}\ \emph {et~al.}(2017)\citenamefont {Gong},
  \citenamefont {Li}, \citenamefont {Li}, \citenamefont {Ji}, \citenamefont
  {Stern}, \citenamefont {Xia}, \citenamefont {Cao}, \citenamefont {Bao},
  \citenamefont {Wang}, \citenamefont {Wang}, \citenamefont {Qiu},
  \citenamefont {Cava}, \citenamefont {Louie}, \citenamefont {Xia},\ and\
  \citenamefont {Zhang}}]{GongZhang2017}%
  \BibitemOpen
  \bibfield  {author} {\bibinfo {author} {\bibfnamefont {C.}~\bibnamefont
  {Gong}}, \bibinfo {author} {\bibfnamefont {L.}~\bibnamefont {Li}}, \bibinfo
  {author} {\bibfnamefont {Z.}~\bibnamefont {Li}}, \bibinfo {author}
  {\bibfnamefont {H.}~\bibnamefont {Ji}}, \bibinfo {author} {\bibfnamefont
  {A.}~\bibnamefont {Stern}}, \bibinfo {author} {\bibfnamefont
  {Y.}~\bibnamefont {Xia}}, \bibinfo {author} {\bibfnamefont {T.}~\bibnamefont
  {Cao}}, \bibinfo {author} {\bibfnamefont {W.}~\bibnamefont {Bao}}, \bibinfo
  {author} {\bibfnamefont {C.}~\bibnamefont {Wang}}, \bibinfo {author}
  {\bibfnamefont {Y.}~\bibnamefont {Wang}}, \bibinfo {author} {\bibfnamefont
  {Z.~Q.}\ \bibnamefont {Qiu}}, \bibinfo {author} {\bibfnamefont {R.~J.}\
  \bibnamefont {Cava}}, \bibinfo {author} {\bibfnamefont {S.~G.}\ \bibnamefont
  {Louie}}, \bibinfo {author} {\bibfnamefont {J.}~\bibnamefont {Xia}},\ and\
  \bibinfo {author} {\bibfnamefont {X.}~\bibnamefont {Zhang}},\ }\bibfield
  {title} {\bibinfo {title} {Discovery of intrinsic ferromagnetism in
  two-dimensional van der waals crystals},\ }\href@noop {} {\bibfield
  {journal} {\bibinfo  {journal} {Nature}\ }\textbf {\bibinfo {volume} {546}},\
  \bibinfo {pages} {265} (\bibinfo {year} {2017})}\BibitemShut {NoStop}%
\bibitem [{\citenamefont {Huang}\ \emph {et~al.}(2017)\citenamefont {Huang},
  \citenamefont {Clark}, \citenamefont {Navarro-Moratalla}, \citenamefont
  {Klein}, \citenamefont {Cheng}, \citenamefont {Seyler}, \citenamefont
  {Zhong}, \citenamefont {Schmidgall}, \citenamefont {McGuire}, \citenamefont
  {Cobden}, \citenamefont {Yao}, \citenamefont {Xiao}, \citenamefont
  {Jarillo-Herrero},\ and\ \citenamefont {Xu}}]{HuangXu2017}%
  \BibitemOpen
  \bibfield  {author} {\bibinfo {author} {\bibfnamefont {B.}~\bibnamefont
  {Huang}}, \bibinfo {author} {\bibfnamefont {G.}~\bibnamefont {Clark}},
  \bibinfo {author} {\bibfnamefont {E.}~\bibnamefont {Navarro-Moratalla}},
  \bibinfo {author} {\bibfnamefont {D.~R.}\ \bibnamefont {Klein}}, \bibinfo
  {author} {\bibfnamefont {R.}~\bibnamefont {Cheng}}, \bibinfo {author}
  {\bibfnamefont {K.~L.}\ \bibnamefont {Seyler}}, \bibinfo {author}
  {\bibfnamefont {D.}~\bibnamefont {Zhong}}, \bibinfo {author} {\bibfnamefont
  {E.}~\bibnamefont {Schmidgall}}, \bibinfo {author} {\bibfnamefont {M.~A.}\
  \bibnamefont {McGuire}}, \bibinfo {author} {\bibfnamefont {D.~H.}\
  \bibnamefont {Cobden}}, \bibinfo {author} {\bibfnamefont {W.}~\bibnamefont
  {Yao}}, \bibinfo {author} {\bibfnamefont {D.}~\bibnamefont {Xiao}}, \bibinfo
  {author} {\bibfnamefont {P.}~\bibnamefont {Jarillo-Herrero}},\ and\ \bibinfo
  {author} {\bibfnamefont {X.}~\bibnamefont {Xu}},\ }\bibfield  {title}
  {\bibinfo {title} {Layer-dependent ferromagnetism in a van der waals crystal
  down to the monolayer limit},\ }\href {https://doi.org/10.1038/nature22391}
  {\bibfield  {journal} {\bibinfo  {journal} {Nature}\ }\textbf {\bibinfo
  {volume} {546}},\ \bibinfo {pages} {270} (\bibinfo {year}
  {2017})}\BibitemShut {NoStop}%
\bibitem [{\citenamefont {Chen}\ \emph {et~al.}(2019)\citenamefont {Chen},
  \citenamefont {Sun}, \citenamefont {Wang}, \citenamefont {Gu}, \citenamefont
  {Xu}, \citenamefont {Wu},\ and\ \citenamefont {Gao}}]{ChenGao2019}%
  \BibitemOpen
  \bibfield  {author} {\bibinfo {author} {\bibfnamefont {W.}~\bibnamefont
  {Chen}}, \bibinfo {author} {\bibfnamefont {Z.}~\bibnamefont {Sun}}, \bibinfo
  {author} {\bibfnamefont {Z.}~\bibnamefont {Wang}}, \bibinfo {author}
  {\bibfnamefont {L.}~\bibnamefont {Gu}}, \bibinfo {author} {\bibfnamefont
  {X.}~\bibnamefont {Xu}}, \bibinfo {author} {\bibfnamefont {S.}~\bibnamefont
  {Wu}},\ and\ \bibinfo {author} {\bibfnamefont {C.}~\bibnamefont {Gao}},\
  }\bibfield  {title} {\bibinfo {title} {Direct observation of van der waals
  stacking dependent interlayer magnetism},\ }\href
  {https://www.science.org/doi/abs/10.1126/science.aav1937} {\bibfield
  {journal} {\bibinfo  {journal} {Science}\ }\textbf {\bibinfo {volume}
  {366}},\ \bibinfo {pages} {983} (\bibinfo {year} {2019})}\BibitemShut
  {NoStop}%
\bibitem [{\citenamefont {Och}\ \emph {et~al.}(2021)\citenamefont {Och},
  \citenamefont {Martin}, \citenamefont {Dlubak}, \citenamefont {Seneor},\ and\
  \citenamefont {Mattevi}}]{OchMattevi2021}%
  \BibitemOpen
  \bibfield  {author} {\bibinfo {author} {\bibfnamefont {M.}~\bibnamefont
  {Och}}, \bibinfo {author} {\bibfnamefont {M.-B.}\ \bibnamefont {Martin}},
  \bibinfo {author} {\bibfnamefont {B.}~\bibnamefont {Dlubak}}, \bibinfo
  {author} {\bibfnamefont {P.}~\bibnamefont {Seneor}},\ and\ \bibinfo {author}
  {\bibfnamefont {C.}~\bibnamefont {Mattevi}},\ }\bibfield  {title} {\bibinfo
  {title} {Synthesis of emerging 2d layered magnetic materials},\ }\href
  {http://dx.doi.org/10.1039/D0NR07867K} {\bibfield  {journal} {\bibinfo
  {journal} {Nanoscale}\ }\textbf {\bibinfo {volume} {13}},\ \bibinfo {pages}
  {2157} (\bibinfo {year} {2021})}\BibitemShut {NoStop}%
\bibitem [{\citenamefont {Gibertini}\ \emph {et~al.}(2019)\citenamefont
  {Gibertini}, \citenamefont {Koperski}, \citenamefont {Morpurgo},\ and\
  \citenamefont {Novoselov}}]{GibertiniNovoselov2019}%
  \BibitemOpen
  \bibfield  {author} {\bibinfo {author} {\bibfnamefont {M.}~\bibnamefont
  {Gibertini}}, \bibinfo {author} {\bibfnamefont {M.}~\bibnamefont {Koperski}},
  \bibinfo {author} {\bibfnamefont {A.~F.}\ \bibnamefont {Morpurgo}},\ and\
  \bibinfo {author} {\bibfnamefont {K.~S.}\ \bibnamefont {Novoselov}},\
  }\bibfield  {title} {\bibinfo {title} {Magnetic 2d materials and
  heterostructures},\ }\href@noop {} {\bibfield  {journal} {\bibinfo  {journal}
  {Nat. Nanotech.}\ }\textbf {\bibinfo {volume} {14}},\ \bibinfo {pages} {408}
  (\bibinfo {year} {2019})}\BibitemShut {NoStop}%
\bibitem [{\citenamefont {Soriano}\ \emph {et~al.}(2019)\citenamefont
  {Soriano}, \citenamefont {Katsnelson},\ and\ \citenamefont
  {Fern\'andez-Rossier}}]{SorianoRossier2020}%
  \BibitemOpen
  \bibfield  {author} {\bibinfo {author} {\bibfnamefont {D.}~\bibnamefont
  {Soriano}}, \bibinfo {author} {\bibfnamefont {M.~I.}\ \bibnamefont
  {Katsnelson}},\ and\ \bibinfo {author} {\bibfnamefont {J.}~\bibnamefont
  {Fern\'andez-Rossier}},\ }\bibfield  {title} {\bibinfo {title} {Magnetic
  two-dimensional chromium trihalides: A theoretical perspective},\ }\href@noop
  {} {\bibfield  {journal} {\bibinfo  {journal} {Nano Letters}\ }\textbf
  {\bibinfo {volume} {20}},\ \bibinfo {pages} {6225} (\bibinfo {year}
  {2019})}\BibitemShut {NoStop}%
\bibitem [{\citenamefont {Klein}\ \emph {et~al.}(2018)\citenamefont {Klein},
  \citenamefont {MacNeill}, \citenamefont {Lado}, \citenamefont {Soriano},
  \citenamefont {Navarro-Moratalla}, \citenamefont {Watanabe}, \citenamefont
  {Taniguchi}, \citenamefont {Manni}, \citenamefont {Canfield}, \citenamefont
  {Fern{\'a}ndez-Rossier},\ and\ \citenamefont
  {Jarillo-Herrero}}]{KleinJarillo2018}%
  \BibitemOpen
  \bibfield  {author} {\bibinfo {author} {\bibfnamefont {D.~R.}\ \bibnamefont
  {Klein}}, \bibinfo {author} {\bibfnamefont {D.}~\bibnamefont {MacNeill}},
  \bibinfo {author} {\bibfnamefont {J.~L.}\ \bibnamefont {Lado}}, \bibinfo
  {author} {\bibfnamefont {D.}~\bibnamefont {Soriano}}, \bibinfo {author}
  {\bibfnamefont {E.}~\bibnamefont {Navarro-Moratalla}}, \bibinfo {author}
  {\bibfnamefont {K.}~\bibnamefont {Watanabe}}, \bibinfo {author}
  {\bibfnamefont {T.}~\bibnamefont {Taniguchi}}, \bibinfo {author}
  {\bibfnamefont {S.}~\bibnamefont {Manni}}, \bibinfo {author} {\bibfnamefont
  {P.}~\bibnamefont {Canfield}}, \bibinfo {author} {\bibfnamefont
  {J.}~\bibnamefont {Fern{\'a}ndez-Rossier}},\ and\ \bibinfo {author}
  {\bibfnamefont {P.}~\bibnamefont {Jarillo-Herrero}},\ }\bibfield  {title}
  {\bibinfo {title} {{Probing magnetism in 2D van der Waals crystalline
  insulators via electron tunneling}},\ }\href
  {https://doi.org/10.1126/science.aar3617} {\bibfield  {journal} {\bibinfo
  {journal} {Science}\ }\textbf {\bibinfo {volume} {360}},\ \bibinfo {pages}
  {1218} (\bibinfo {year} {2018})}\BibitemShut {NoStop}%
\bibitem [{\citenamefont {Ghazaryan}\ \emph {et~al.}(2018)\citenamefont
  {Ghazaryan}, \citenamefont {Greenaway}, \citenamefont {Wang}, \citenamefont
  {Guarochico-Moreira}, \citenamefont {Vera-Marun}, \citenamefont {Yin},
  \citenamefont {Liao}, \citenamefont {Morozov}, \citenamefont {Kristanovski},
  \citenamefont {Lichtenstein}, \citenamefont {Katsnelson}, \citenamefont
  {Withers}, \citenamefont {Mishchenko}, \citenamefont {Eaves}, \citenamefont
  {Geim}, \citenamefont {Novoselov},\ and\ \citenamefont
  {Misra}}]{GhazaryanMisra2018}%
  \BibitemOpen
  \bibfield  {author} {\bibinfo {author} {\bibfnamefont {D.}~\bibnamefont
  {Ghazaryan}}, \bibinfo {author} {\bibfnamefont {M.~T.}\ \bibnamefont
  {Greenaway}}, \bibinfo {author} {\bibfnamefont {Z.}~\bibnamefont {Wang}},
  \bibinfo {author} {\bibfnamefont {V.~H.}\ \bibnamefont {Guarochico-Moreira}},
  \bibinfo {author} {\bibfnamefont {I.~J.}\ \bibnamefont {Vera-Marun}},
  \bibinfo {author} {\bibfnamefont {J.}~\bibnamefont {Yin}}, \bibinfo {author}
  {\bibfnamefont {Y.}~\bibnamefont {Liao}}, \bibinfo {author} {\bibfnamefont
  {S.~V.}\ \bibnamefont {Morozov}}, \bibinfo {author} {\bibfnamefont
  {O.}~\bibnamefont {Kristanovski}}, \bibinfo {author} {\bibfnamefont {A.~I.}\
  \bibnamefont {Lichtenstein}}, \bibinfo {author} {\bibfnamefont {M.~I.}\
  \bibnamefont {Katsnelson}}, \bibinfo {author} {\bibfnamefont
  {F.}~\bibnamefont {Withers}}, \bibinfo {author} {\bibfnamefont
  {A.}~\bibnamefont {Mishchenko}}, \bibinfo {author} {\bibfnamefont
  {L.}~\bibnamefont {Eaves}}, \bibinfo {author} {\bibfnamefont {A.~K.}\
  \bibnamefont {Geim}}, \bibinfo {author} {\bibfnamefont {K.~S.}\ \bibnamefont
  {Novoselov}},\ and\ \bibinfo {author} {\bibfnamefont {A.}~\bibnamefont
  {Misra}},\ }\bibfield  {title} {\bibinfo {title} {{Magnon-assisted tunnelling
  in van der Waals heterostructures based on CrBr$_3$}},\ }\href
  {https://doi.org/10.1038/s41928-018-0087-z} {\bibfield  {journal} {\bibinfo
  {journal} {Nat. Electron.}\ }\textbf {\bibinfo {volume} {1}},\ \bibinfo
  {pages} {344} (\bibinfo {year} {2018})}\BibitemShut {NoStop}%
\bibitem [{\citenamefont {Wang}\ \emph {et~al.}(2018)\citenamefont {Wang},
  \citenamefont {Guti{\'e}rrez-Lezama}, \citenamefont {Ubrig}, \citenamefont
  {Kroner}, \citenamefont {Gibertini}, \citenamefont {Taniguchi}, \citenamefont
  {Watanabe}, \citenamefont {Imamo{\u g}lu}, \citenamefont {Giannini},\ and\
  \citenamefont {Morpurgo}}]{WangMorpurgo2018}%
  \BibitemOpen
  \bibfield  {author} {\bibinfo {author} {\bibfnamefont {Z.}~\bibnamefont
  {Wang}}, \bibinfo {author} {\bibfnamefont {I.}~\bibnamefont
  {Guti{\'e}rrez-Lezama}}, \bibinfo {author} {\bibfnamefont {N.}~\bibnamefont
  {Ubrig}}, \bibinfo {author} {\bibfnamefont {M.}~\bibnamefont {Kroner}},
  \bibinfo {author} {\bibfnamefont {M.}~\bibnamefont {Gibertini}}, \bibinfo
  {author} {\bibfnamefont {T.}~\bibnamefont {Taniguchi}}, \bibinfo {author}
  {\bibfnamefont {K.}~\bibnamefont {Watanabe}}, \bibinfo {author}
  {\bibfnamefont {A.}~\bibnamefont {Imamo{\u g}lu}}, \bibinfo {author}
  {\bibfnamefont {E.}~\bibnamefont {Giannini}},\ and\ \bibinfo {author}
  {\bibfnamefont {A.~F.}\ \bibnamefont {Morpurgo}},\ }\bibfield  {title}
  {\bibinfo {title} {Very large tunneling magnetoresistance in layered magnetic
  semiconductor cri$_3$},\ }\href@noop {} {\bibfield  {journal} {\bibinfo
  {journal} {Nature Communications}\ }\textbf {\bibinfo {volume} {9}},\
  \bibinfo {pages} {2516} (\bibinfo {year} {2018})}\BibitemShut {NoStop}%
\bibitem [{\citenamefont {Ai}\ \emph {et~al.}(2021)\citenamefont {Ai},
  \citenamefont {Zhang}, \citenamefont {Yang}, \citenamefont {Xie},
  \citenamefont {Yang}, \citenamefont {Jia}, \citenamefont {Zhang},
  \citenamefont {Liu}, \citenamefont {Li}, \citenamefont {Leng}, \citenamefont
  {Cao}, \citenamefont {Sun}, \citenamefont {Zhang}, \citenamefont {Kou},
  \citenamefont {Han}, \citenamefont {Xiu},\ and\ \citenamefont
  {Dong}}]{AiDong2021}%
  \BibitemOpen
  \bibfield  {author} {\bibinfo {author} {\bibfnamefont {L.}~\bibnamefont
  {Ai}}, \bibinfo {author} {\bibfnamefont {E.}~\bibnamefont {Zhang}}, \bibinfo
  {author} {\bibfnamefont {J.}~\bibnamefont {Yang}}, \bibinfo {author}
  {\bibfnamefont {X.}~\bibnamefont {Xie}}, \bibinfo {author} {\bibfnamefont
  {Y.}~\bibnamefont {Yang}}, \bibinfo {author} {\bibfnamefont {Z.}~\bibnamefont
  {Jia}}, \bibinfo {author} {\bibfnamefont {Y.}~\bibnamefont {Zhang}}, \bibinfo
  {author} {\bibfnamefont {S.}~\bibnamefont {Liu}}, \bibinfo {author}
  {\bibfnamefont {Z.}~\bibnamefont {Li}}, \bibinfo {author} {\bibfnamefont
  {P.}~\bibnamefont {Leng}}, \bibinfo {author} {\bibfnamefont {X.}~\bibnamefont
  {Cao}}, \bibinfo {author} {\bibfnamefont {X.}~\bibnamefont {Sun}}, \bibinfo
  {author} {\bibfnamefont {T.}~\bibnamefont {Zhang}}, \bibinfo {author}
  {\bibfnamefont {X.}~\bibnamefont {Kou}}, \bibinfo {author} {\bibfnamefont
  {Z.}~\bibnamefont {Han}}, \bibinfo {author} {\bibfnamefont {F.}~\bibnamefont
  {Xiu}},\ and\ \bibinfo {author} {\bibfnamefont {S.}~\bibnamefont {Dong}},\
  }\bibfield  {title} {\bibinfo {title} {Van der waals ferromagnetic josephson
  junctions},\ }\href {https://doi.org/10.1038/s41467-021-26946-w} {\bibfield
  {journal} {\bibinfo  {journal} {Nature Communications}\ }\textbf {\bibinfo
  {volume} {12}},\ \bibinfo {pages} {6580} (\bibinfo {year}
  {2021})}\BibitemShut {NoStop}%
\bibitem [{\citenamefont {Wu}\ \emph {et~al.}(2020)\citenamefont {Wu},
  \citenamefont {Zhang}, \citenamefont {Zhang}, \citenamefont {Wang},
  \citenamefont {Zhu}, \citenamefont {Hu}, \citenamefont {Yin}, \citenamefont
  {Wong}, \citenamefont {Fang}, \citenamefont {Wan}, \citenamefont {Han},
  \citenamefont {Shao}, \citenamefont {Taniguchi}, \citenamefont {Watanabe},
  \citenamefont {Zang}, \citenamefont {Mao}, \citenamefont {Zhang},\ and\
  \citenamefont {Wang}}]{WuKang2020}%
  \BibitemOpen
  \bibfield  {author} {\bibinfo {author} {\bibfnamefont {Y.}~\bibnamefont
  {Wu}}, \bibinfo {author} {\bibfnamefont {S.}~\bibnamefont {Zhang}}, \bibinfo
  {author} {\bibfnamefont {J.}~\bibnamefont {Zhang}}, \bibinfo {author}
  {\bibfnamefont {W.}~\bibnamefont {Wang}}, \bibinfo {author} {\bibfnamefont
  {Y.~L.}\ \bibnamefont {Zhu}}, \bibinfo {author} {\bibfnamefont
  {J.}~\bibnamefont {Hu}}, \bibinfo {author} {\bibfnamefont {G.}~\bibnamefont
  {Yin}}, \bibinfo {author} {\bibfnamefont {K.}~\bibnamefont {Wong}}, \bibinfo
  {author} {\bibfnamefont {C.}~\bibnamefont {Fang}}, \bibinfo {author}
  {\bibfnamefont {C.}~\bibnamefont {Wan}}, \bibinfo {author} {\bibfnamefont
  {X.}~\bibnamefont {Han}}, \bibinfo {author} {\bibfnamefont {Q.}~\bibnamefont
  {Shao}}, \bibinfo {author} {\bibfnamefont {T.}~\bibnamefont {Taniguchi}},
  \bibinfo {author} {\bibfnamefont {K.}~\bibnamefont {Watanabe}}, \bibinfo
  {author} {\bibfnamefont {J.}~\bibnamefont {Zang}}, \bibinfo {author}
  {\bibfnamefont {Z.}~\bibnamefont {Mao}}, \bibinfo {author} {\bibfnamefont
  {X.}~\bibnamefont {Zhang}},\ and\ \bibinfo {author} {\bibfnamefont {K.~L.}\
  \bibnamefont {Wang}},\ }\bibfield  {title} {\bibinfo {title} {N{\'e}el-type
  skyrmion in wte2/fe3gete2 van der waals heterostructure},\ }\href
  {https://doi.org/10.1038/s41467-020-17566-x} {\bibfield  {journal} {\bibinfo
  {journal} {Nature Communications}\ }\textbf {\bibinfo {volume} {11}},\
  \bibinfo {pages} {3860} (\bibinfo {year} {2020})}\BibitemShut {NoStop}%
\bibitem [{\citenamefont {Zhong}\ \emph {et~al.}(2017)\citenamefont {Zhong},
  \citenamefont {Seyler}, \citenamefont {Linpeng}, \citenamefont {Cheng},
  \citenamefont {Sivadas}, \citenamefont {Huang}, \citenamefont {Schmidgall},
  \citenamefont {Taniguchi}, \citenamefont {Watanabe}, \citenamefont {McGuire},
  \citenamefont {Yao}, \citenamefont {Xiao}, \citenamefont {Fu},\ and\
  \citenamefont {Xu}}]{ZhongXu2017}%
  \BibitemOpen
  \bibfield  {author} {\bibinfo {author} {\bibfnamefont {D.}~\bibnamefont
  {Zhong}}, \bibinfo {author} {\bibfnamefont {K.~L.}\ \bibnamefont {Seyler}},
  \bibinfo {author} {\bibfnamefont {X.}~\bibnamefont {Linpeng}}, \bibinfo
  {author} {\bibfnamefont {R.}~\bibnamefont {Cheng}}, \bibinfo {author}
  {\bibfnamefont {N.}~\bibnamefont {Sivadas}}, \bibinfo {author} {\bibfnamefont
  {B.}~\bibnamefont {Huang}}, \bibinfo {author} {\bibfnamefont
  {E.}~\bibnamefont {Schmidgall}}, \bibinfo {author} {\bibfnamefont
  {T.}~\bibnamefont {Taniguchi}}, \bibinfo {author} {\bibfnamefont
  {K.}~\bibnamefont {Watanabe}}, \bibinfo {author} {\bibfnamefont {M.~A.}\
  \bibnamefont {McGuire}}, \bibinfo {author} {\bibfnamefont {W.}~\bibnamefont
  {Yao}}, \bibinfo {author} {\bibfnamefont {D.}~\bibnamefont {Xiao}}, \bibinfo
  {author} {\bibfnamefont {K.-M.~C.}\ \bibnamefont {Fu}},\ and\ \bibinfo
  {author} {\bibfnamefont {X.}~\bibnamefont {Xu}},\ }\bibfield  {title}
  {\bibinfo {title} {Van der waals engineering of ferromagnetic semiconductor
  heterostructures for spin and valleytronics},\ }\href@noop {} {\bibfield
  {journal} {\bibinfo  {journal} {Science Advances}\ }\textbf {\bibinfo
  {volume} {3}},\ \bibinfo {pages} {e1603113} (\bibinfo {year}
  {2017})}\BibitemShut {NoStop}%
\bibitem [{\citenamefont {Zhang}\ \emph {et~al.}(2019)\citenamefont {Zhang},
  \citenamefont {Ni}, \citenamefont {Huang}, \citenamefont {Hu},\ and\
  \citenamefont {Liu}}]{ZhangLiu2019}%
  \BibitemOpen
  \bibfield  {author} {\bibinfo {author} {\bibfnamefont {Z.}~\bibnamefont
  {Zhang}}, \bibinfo {author} {\bibfnamefont {X.}~\bibnamefont {Ni}}, \bibinfo
  {author} {\bibfnamefont {H.}~\bibnamefont {Huang}}, \bibinfo {author}
  {\bibfnamefont {L.}~\bibnamefont {Hu}},\ and\ \bibinfo {author}
  {\bibfnamefont {F.}~\bibnamefont {Liu}},\ }\bibfield  {title} {\bibinfo
  {title} {Valley splitting in the van der waals heterostructure
  ${\mathrm{wse}}_{2}/{\mathrm{cri}}_{3}$: The role of atom superposition},\
  }\href {https://doi.org/10.1103/PhysRevB.99.115441} {\bibfield  {journal}
  {\bibinfo  {journal} {Phys. Rev. B}\ }\textbf {\bibinfo {volume} {99}},\
  \bibinfo {pages} {115441} (\bibinfo {year} {2019})}\BibitemShut {NoStop}%
\bibitem [{\citenamefont {Ge}\ \emph {et~al.}(2022)\citenamefont {Ge},
  \citenamefont {Wang}, \citenamefont {Wu}, \citenamefont {Si}, \citenamefont
  {Zhang},\ and\ \citenamefont {Zhang}}]{GeZhang2022}%
  \BibitemOpen
  \bibfield  {author} {\bibinfo {author} {\bibfnamefont {M.}~\bibnamefont
  {Ge}}, \bibinfo {author} {\bibfnamefont {H.}~\bibnamefont {Wang}}, \bibinfo
  {author} {\bibfnamefont {J.}~\bibnamefont {Wu}}, \bibinfo {author}
  {\bibfnamefont {C.}~\bibnamefont {Si}}, \bibinfo {author} {\bibfnamefont
  {J.}~\bibnamefont {Zhang}},\ and\ \bibinfo {author} {\bibfnamefont
  {S.}~\bibnamefont {Zhang}},\ }\bibfield  {title} {\bibinfo {title} {Enhanced
  valley splitting of wse2 in twisted van der waals wse2/cri3
  heterostructures},\ }\href {https://doi.org/10.1038/s41524-022-00715-9}
  {\bibfield  {journal} {\bibinfo  {journal} {npj Computational Materials}\
  }\textbf {\bibinfo {volume} {8}},\ \bibinfo {pages} {32} (\bibinfo {year}
  {2022})}\BibitemShut {NoStop}%
\bibitem [{\citenamefont {Zollner}\ \emph {et~al.}(2023)\citenamefont
  {Zollner}, \citenamefont {Faria~Junior},\ and\ \citenamefont
  {Fabian}}]{ZollnerFabian2023}%
  \BibitemOpen
  \bibfield  {author} {\bibinfo {author} {\bibfnamefont {K.}~\bibnamefont
  {Zollner}}, \bibinfo {author} {\bibfnamefont {P.~E.}\ \bibnamefont
  {Faria~Junior}},\ and\ \bibinfo {author} {\bibfnamefont {J.}~\bibnamefont
  {Fabian}},\ }\bibfield  {title} {\bibinfo {title} {Strong manipulation of the
  valley splitting upon twisting and gating in
  ${\mathrm{mose}}_{2}/{\mathrm{cri}}_{3}$ and
  ${\mathrm{wse}}_{2}/{\mathrm{cri}}_{3}$ van der waals heterostructures},\
  }\href {https://doi.org/10.1103/PhysRevB.107.035112} {\bibfield  {journal}
  {\bibinfo  {journal} {Phys. Rev. B}\ }\textbf {\bibinfo {volume} {107}},\
  \bibinfo {pages} {035112} (\bibinfo {year} {2023})}\BibitemShut {NoStop}%
\bibitem [{\citenamefont {Ciorciaro}\ \emph {et~al.}(2020)\citenamefont
  {Ciorciaro}, \citenamefont {Kroner}, \citenamefont {Watanabe}, \citenamefont
  {Taniguchi},\ and\ \citenamefont {Imamoglu}}]{CiorciaroImamoglu2020}%
  \BibitemOpen
  \bibfield  {author} {\bibinfo {author} {\bibfnamefont {L.}~\bibnamefont
  {Ciorciaro}}, \bibinfo {author} {\bibfnamefont {M.}~\bibnamefont {Kroner}},
  \bibinfo {author} {\bibfnamefont {K.}~\bibnamefont {Watanabe}}, \bibinfo
  {author} {\bibfnamefont {T.}~\bibnamefont {Taniguchi}},\ and\ \bibinfo
  {author} {\bibfnamefont {A.}~\bibnamefont {Imamoglu}},\ }\bibfield  {title}
  {\bibinfo {title} {Observation of magnetic proximity effect using resonant
  optical spectroscopy of an electrically tunable
  ${\mathrm{mose}}_{2}/{\mathrm{crbr}}_{3}$ heterostructure},\ }\href
  {https://link.aps.org/doi/10.1103/PhysRevLett.124.197401} {\bibfield
  {journal} {\bibinfo  {journal} {Phys. Rev. Lett.}\ }\textbf {\bibinfo
  {volume} {124}},\ \bibinfo {pages} {197401} (\bibinfo {year}
  {2020})}\BibitemShut {NoStop}%
\bibitem [{\citenamefont {Choi}\ \emph {et~al.}(2023)\citenamefont {Choi},
  \citenamefont {Lane}, \citenamefont {Zhu},\ and\ \citenamefont
  {Crooker}}]{ChoiCrooker2023}%
  \BibitemOpen
  \bibfield  {author} {\bibinfo {author} {\bibfnamefont {J.}~\bibnamefont
  {Choi}}, \bibinfo {author} {\bibfnamefont {C.}~\bibnamefont {Lane}}, \bibinfo
  {author} {\bibfnamefont {J.-X.}\ \bibnamefont {Zhu}},\ and\ \bibinfo {author}
  {\bibfnamefont {S.~A.}\ \bibnamefont {Crooker}},\ }\bibfield  {title}
  {\bibinfo {title} {Asymmetric magnetic proximity interactions in mose2/crbr3
  van der waals heterostructures},\ }\href@noop {} {\bibfield  {journal}
  {\bibinfo  {journal} {Nature Materials}\ }\textbf {\bibinfo {volume} {22}},\
  \bibinfo {pages} {305} (\bibinfo {year} {2023})}\BibitemShut {NoStop}%
\bibitem [{\citenamefont {Jiang}\ \emph
  {et~al.}(2018{\natexlab{a}})\citenamefont {Jiang}, \citenamefont {Li},
  \citenamefont {Wang}, \citenamefont {Mak},\ and\ \citenamefont
  {Shan}}]{JiangShan2018}%
  \BibitemOpen
  \bibfield  {author} {\bibinfo {author} {\bibfnamefont {S.}~\bibnamefont
  {Jiang}}, \bibinfo {author} {\bibfnamefont {L.}~\bibnamefont {Li}}, \bibinfo
  {author} {\bibfnamefont {Z.}~\bibnamefont {Wang}}, \bibinfo {author}
  {\bibfnamefont {K.~F.}\ \bibnamefont {Mak}},\ and\ \bibinfo {author}
  {\bibfnamefont {J.}~\bibnamefont {Shan}},\ }\bibfield  {title} {\bibinfo
  {title} {{Controlling magnetism in 2D CrI$_3$ by electrostatic doping}},\
  }\href {https://doi.org/10.1038/s41565-018-0135-x} {\bibfield  {journal}
  {\bibinfo  {journal} {Nat. Nanotech.}\ }\textbf {\bibinfo {volume} {13}},\
  \bibinfo {pages} {549} (\bibinfo {year} {2018}{\natexlab{a}})}\BibitemShut
  {NoStop}%
\bibitem [{\citenamefont {Jiang}\ \emph
  {et~al.}(2018{\natexlab{b}})\citenamefont {Jiang}, \citenamefont {Shan},\
  and\ \citenamefont {Mak}}]{JiangMak2018}%
  \BibitemOpen
  \bibfield  {author} {\bibinfo {author} {\bibfnamefont {S.}~\bibnamefont
  {Jiang}}, \bibinfo {author} {\bibfnamefont {J.}~\bibnamefont {Shan}},\ and\
  \bibinfo {author} {\bibfnamefont {K.~F.}\ \bibnamefont {Mak}},\ }\bibfield
  {title} {\bibinfo {title} {Electric-field switching of two-dimensional van
  der waals magnets},\ }\href
  {https://doi.org/doi.org/10.1038/s41563-018-0040-6} {\bibfield  {journal}
  {\bibinfo  {journal} {Nat. Mater.}\ }\textbf {\bibinfo {volume} {17}},\
  \bibinfo {pages} {406} (\bibinfo {year} {2018}{\natexlab{b}})}\BibitemShut
  {NoStop}%
\bibitem [{\citenamefont {Huang}\ \emph {et~al.}(2018)\citenamefont {Huang},
  \citenamefont {Clark}, \citenamefont {Klein}, \citenamefont {MacNeill},
  \citenamefont {Navarro-Moratalla}, \citenamefont {Seyler}, \citenamefont
  {Wilson}, \citenamefont {McGuire}, \citenamefont {Cobden}, \citenamefont
  {Xiao}, \citenamefont {Yao}, \citenamefont {Jarillo-Herrero},\ and\
  \citenamefont {Xu}}]{HuangXu2018}%
  \BibitemOpen
  \bibfield  {author} {\bibinfo {author} {\bibfnamefont {B.}~\bibnamefont
  {Huang}}, \bibinfo {author} {\bibfnamefont {G.}~\bibnamefont {Clark}},
  \bibinfo {author} {\bibfnamefont {D.~R.}\ \bibnamefont {Klein}}, \bibinfo
  {author} {\bibfnamefont {D.}~\bibnamefont {MacNeill}}, \bibinfo {author}
  {\bibfnamefont {E.}~\bibnamefont {Navarro-Moratalla}}, \bibinfo {author}
  {\bibfnamefont {K.~L.}\ \bibnamefont {Seyler}}, \bibinfo {author}
  {\bibfnamefont {N.}~\bibnamefont {Wilson}}, \bibinfo {author} {\bibfnamefont
  {M.~A.}\ \bibnamefont {McGuire}}, \bibinfo {author} {\bibfnamefont {D.~H.}\
  \bibnamefont {Cobden}}, \bibinfo {author} {\bibfnamefont {D.}~\bibnamefont
  {Xiao}}, \bibinfo {author} {\bibfnamefont {W.}~\bibnamefont {Yao}}, \bibinfo
  {author} {\bibfnamefont {P.}~\bibnamefont {Jarillo-Herrero}},\ and\ \bibinfo
  {author} {\bibfnamefont {X.}~\bibnamefont {Xu}},\ }\bibfield  {title}
  {\bibinfo {title} {{Electrical control of 2D magnetism in bilayer CrI$_3$}},\
  }\href {https://doi.org/10.1038/s41565-018-0121-3} {\bibfield  {journal}
  {\bibinfo  {journal} {Nat. Nanotech.}\ }\textbf {\bibinfo {volume} {13}},\
  \bibinfo {pages} {544} (\bibinfo {year} {2018})}\BibitemShut {NoStop}%
\bibitem [{\citenamefont {Liu}\ \emph {et~al.}(2013)\citenamefont {Liu},
  \citenamefont {Kang}, \citenamefont {Sarkar}, \citenamefont {Khatami},
  \citenamefont {Jena},\ and\ \citenamefont {Banerjee}}]{Liu2013}%
  \BibitemOpen
  \bibfield  {author} {\bibinfo {author} {\bibfnamefont {W.}~\bibnamefont
  {Liu}}, \bibinfo {author} {\bibfnamefont {J.}~\bibnamefont {Kang}}, \bibinfo
  {author} {\bibfnamefont {D.}~\bibnamefont {Sarkar}}, \bibinfo {author}
  {\bibfnamefont {Y.}~\bibnamefont {Khatami}}, \bibinfo {author} {\bibfnamefont
  {D.}~\bibnamefont {Jena}},\ and\ \bibinfo {author} {\bibfnamefont
  {K.}~\bibnamefont {Banerjee}},\ }\bibfield  {title} {\bibinfo {title} {Role
  of metal contacts in designing high-performance monolayer n-type wse2 field
  effect transistors},\ }\href {https://doi.org/10.1021/nl304777e} {\bibfield
  {journal} {\bibinfo  {journal} {Nano Letters}\ }\textbf {\bibinfo {volume}
  {13}},\ \bibinfo {pages} {1983} (\bibinfo {year} {2013})},\ \bibinfo {note}
  {pMID: 23527483},\ \Eprint
  {https://arxiv.org/abs/https://doi.org/10.1021/nl304777e}
  {https://doi.org/10.1021/nl304777e} \BibitemShut {NoStop}%
\bibitem [{\citenamefont {Deng}\ \emph {et~al.}(2018)\citenamefont {Deng},
  \citenamefont {Li},\ and\ \citenamefont {Li}}]{DENG201844}%
  \BibitemOpen
  \bibfield  {author} {\bibinfo {author} {\bibfnamefont {S.}~\bibnamefont
  {Deng}}, \bibinfo {author} {\bibfnamefont {L.}~\bibnamefont {Li}},\ and\
  \bibinfo {author} {\bibfnamefont {M.}~\bibnamefont {Li}},\ }\bibfield
  {title} {\bibinfo {title} {Stability of direct band gap under mechanical
  strains for monolayer mos2, mose2, ws2 and wse2},\ }\href
  {https://doi.org/https://doi.org/10.1016/j.physe.2018.03.016} {\bibfield
  {journal} {\bibinfo  {journal} {Physica E: Low-dimensional Systems and
  Nanostructures}\ }\textbf {\bibinfo {volume} {101}},\ \bibinfo {pages} {44}
  (\bibinfo {year} {2018})}\BibitemShut {NoStop}%
\bibitem [{\citenamefont {Sohier}\ \emph {et~al.}(2023)\citenamefont {Sohier},
  \citenamefont {de~Melo}, \citenamefont {Zanolli},\ and\ \citenamefont
  {Verstraete}}]{SohierVerstraete2023}%
  \BibitemOpen
  \bibfield  {author} {\bibinfo {author} {\bibfnamefont {T.}~\bibnamefont
  {Sohier}}, \bibinfo {author} {\bibfnamefont {P.~M. M.~C.}\ \bibnamefont
  {de~Melo}}, \bibinfo {author} {\bibfnamefont {Z.}~\bibnamefont {Zanolli}},\
  and\ \bibinfo {author} {\bibfnamefont {M.~J.}\ \bibnamefont {Verstraete}},\
  }\bibfield  {title} {\bibinfo {title} {The impact of valley profile on the
  mobility and kerr rotation of transition metal dichalcogenides},\ }\href
  {https://doi.org/10.1088/2053-1583/acb21c} {\bibfield  {journal} {\bibinfo
  {journal} {2D Mater.}\ }\textbf {\bibinfo {volume} {10}},\ \bibinfo {pages}
  {025006} (\bibinfo {year} {2023})}\BibitemShut {NoStop}%
\bibitem [{\citenamefont {{\it et al.}}(2009)}]{QE}%
  \BibitemOpen
  \bibfield  {author} {\bibinfo {author} {\bibfnamefont {P.~G.}\ \bibnamefont
  {{\it et al.}}},\ }\bibfield  {title} {\bibinfo {title} {Quantum espresso: a
  modular and open-source software project for quantum simulations of
  materials},\ }\href@noop {} {\bibfield  {journal} {\bibinfo  {journal}
  {Journal of Physics: Condensed Matter}\ }\textbf {\bibinfo {volume} {21}},\
  \bibinfo {pages} {395502} (\bibinfo {year} {2009})}\BibitemShut {NoStop}%
\bibitem [{\citenamefont {Soriano}\ \emph {et~al.}(2021)\citenamefont
  {Soriano}, \citenamefont {Rudenko}, \citenamefont {Katsnelson},\ and\
  \citenamefont {R{\"o}sner}}]{SorianoRosner2021}%
  \BibitemOpen
  \bibfield  {author} {\bibinfo {author} {\bibfnamefont {D.}~\bibnamefont
  {Soriano}}, \bibinfo {author} {\bibfnamefont {A.~N.}\ \bibnamefont
  {Rudenko}}, \bibinfo {author} {\bibfnamefont {M.~I.}\ \bibnamefont
  {Katsnelson}},\ and\ \bibinfo {author} {\bibfnamefont {M.}~\bibnamefont
  {R{\"o}sner}},\ }\bibfield  {title} {\bibinfo {title} {Environmental
  screening and ligand-field effects to magnetism in cri3 monolayer},\ }\href
  {https://doi.org/10.1038/s41524-021-00631-4} {\bibfield  {journal} {\bibinfo
  {journal} {npj Computational Materials}\ }\textbf {\bibinfo {volume} {7}},\
  \bibinfo {pages} {162} (\bibinfo {year} {2021})}\BibitemShut {NoStop}%
\bibitem [{\citenamefont {Mostofi}\ \emph {et~al.}(2008)\citenamefont
  {Mostofi}, \citenamefont {Yates}, \citenamefont {Lee}, \citenamefont {Souza},
  \citenamefont {Vanderbilt},\ and\ \citenamefont {Marzari}}]{Wannier90}%
  \BibitemOpen
  \bibfield  {author} {\bibinfo {author} {\bibfnamefont {A.~A.}\ \bibnamefont
  {Mostofi}}, \bibinfo {author} {\bibfnamefont {J.~R.}\ \bibnamefont {Yates}},
  \bibinfo {author} {\bibfnamefont {Y.-S.}\ \bibnamefont {Lee}}, \bibinfo
  {author} {\bibfnamefont {I.}~\bibnamefont {Souza}}, \bibinfo {author}
  {\bibfnamefont {D.}~\bibnamefont {Vanderbilt}},\ and\ \bibinfo {author}
  {\bibfnamefont {N.}~\bibnamefont {Marzari}},\ }\bibfield  {title} {\bibinfo
  {title} {{wannier90: A tool for obtaining maximally-localised Wannier
  functions}},\ }\href {https://doi.org/doi.org/10.1016/j.cpc.2007.11.016}
  {\bibfield  {journal} {\bibinfo  {journal} {Comp. Phys. Commun.}\ }\textbf
  {\bibinfo {volume} {178}},\ \bibinfo {pages} {685 } (\bibinfo {year}
  {2008})}\BibitemShut {NoStop}%
\bibitem [{\citenamefont {Cannav\`o}\ \emph {et~al.}(2021)\citenamefont
  {Cannav\`o}, \citenamefont {Marian}, \citenamefont {Mar\'{\i}n},
  \citenamefont {Iannaccone},\ and\ \citenamefont {Fiori}}]{Cannavo2021}%
  \BibitemOpen
  \bibfield  {author} {\bibinfo {author} {\bibfnamefont {E.}~\bibnamefont
  {Cannav\`o}}, \bibinfo {author} {\bibfnamefont {D.}~\bibnamefont {Marian}},
  \bibinfo {author} {\bibfnamefont {E.~G.}\ \bibnamefont {Mar\'{\i}n}},
  \bibinfo {author} {\bibfnamefont {G.}~\bibnamefont {Iannaccone}},\ and\
  \bibinfo {author} {\bibfnamefont {G.}~\bibnamefont {Fiori}},\ }\bibfield
  {title} {\bibinfo {title} {Transport properties in partially overlapping van
  der waals junctions through a multiscale investigation},\ }\href
  {https://doi.org/10.1103/PhysRevB.104.085433} {\bibfield  {journal} {\bibinfo
   {journal} {Phys. Rev. B}\ }\textbf {\bibinfo {volume} {104}},\ \bibinfo
  {pages} {085433} (\bibinfo {year} {2021})}\BibitemShut {NoStop}%
\bibitem [{\citenamefont {Katagiri}\ \emph {et~al.}(2016)\citenamefont
  {Katagiri}, \citenamefont {Nakamura}, \citenamefont {Ishii}, \citenamefont
  {Ohata}, \citenamefont {Hasegawa}, \citenamefont {Katsumoto}, \citenamefont
  {Cusati}, \citenamefont {Fortunelli}, \citenamefont {Iannaccone},
  \citenamefont {Fiori}, \citenamefont {Roche},\ and\ \citenamefont
  {Haruyama}}]{Katagiri2016}%
  \BibitemOpen
  \bibfield  {author} {\bibinfo {author} {\bibfnamefont {Y.}~\bibnamefont
  {Katagiri}}, \bibinfo {author} {\bibfnamefont {T.}~\bibnamefont {Nakamura}},
  \bibinfo {author} {\bibfnamefont {A.}~\bibnamefont {Ishii}}, \bibinfo
  {author} {\bibfnamefont {C.}~\bibnamefont {Ohata}}, \bibinfo {author}
  {\bibfnamefont {M.}~\bibnamefont {Hasegawa}}, \bibinfo {author}
  {\bibfnamefont {S.}~\bibnamefont {Katsumoto}}, \bibinfo {author}
  {\bibfnamefont {T.}~\bibnamefont {Cusati}}, \bibinfo {author} {\bibfnamefont
  {A.}~\bibnamefont {Fortunelli}}, \bibinfo {author} {\bibfnamefont
  {G.}~\bibnamefont {Iannaccone}}, \bibinfo {author} {\bibfnamefont
  {G.}~\bibnamefont {Fiori}}, \bibinfo {author} {\bibfnamefont
  {S.}~\bibnamefont {Roche}},\ and\ \bibinfo {author} {\bibfnamefont
  {J.}~\bibnamefont {Haruyama}},\ }\bibfield  {title} {\bibinfo {title}
  {Gate-tunable atomically thin lateral mos2 schottky junction patterned by
  electron beam},\ }\href {https://doi.org/10.1021/acs.nanolett.6b01186}
  {\bibfield  {journal} {\bibinfo  {journal} {Nano Letters}\ }\textbf {\bibinfo
  {volume} {16}},\ \bibinfo {pages} {3788} (\bibinfo {year} {2016})},\ \bibinfo
  {note} {pMID: 27152475},\ \Eprint
  {https://arxiv.org/abs/https://doi.org/10.1021/acs.nanolett.6b01186}
  {https://doi.org/10.1021/acs.nanolett.6b01186} \BibitemShut {NoStop}%
\bibitem [{vid()}]{vides}%
  \BibitemOpen
  \href@noop {} {\bibinfo {title} {{NanoTCAD ViDES}}},\ \bibinfo {howpublished}
  {Available at: {http://vides.nanotcad.com/vides/}}\BibitemShut {NoStop}%
\bibitem [{\citenamefont {Marian}\ \emph {et~al.}(2023)\citenamefont {Marian},
  \citenamefont {Marin}, \citenamefont {Perucchini}, \citenamefont
  {Iannaccone},\ and\ \citenamefont {Fiori}}]{Vides2023}%
  \BibitemOpen
  \bibfield  {author} {\bibinfo {author} {\bibfnamefont {D.}~\bibnamefont
  {Marian}}, \bibinfo {author} {\bibfnamefont {E.~G.}\ \bibnamefont {Marin}},
  \bibinfo {author} {\bibfnamefont {M.}~\bibnamefont {Perucchini}}, \bibinfo
  {author} {\bibfnamefont {G.}~\bibnamefont {Iannaccone}},\ and\ \bibinfo
  {author} {\bibfnamefont {G.}~\bibnamefont {Fiori}},\ }\bibfield  {title}
  {\bibinfo {title} {Multi-scale simulations of two dimensional material based
  devices: the nanotcad vides suite},\ }\bibfield  {journal} {\bibinfo
  {journal} {Journal of Computational Electronics}\ }\href
  {https://doi.org/10.1007/s10825-023-02048-2} {10.1007/s10825-023-02048-2}
  (\bibinfo {year} {2023}),\ \bibinfo {note} {cited by: 1; All Open Access,
  Hybrid Gold Open Access}\BibitemShut {NoStop}%
\bibitem [{\citenamefont {Ko\ifmmode~\acute{s}\else \'{s}\fi{}mider}\ \emph
  {et~al.}(2013)\citenamefont {Ko\ifmmode~\acute{s}\else \'{s}\fi{}mider},
  \citenamefont {Gonz\'alez},\ and\ \citenamefont
  {Fern\'andez-Rossier}}]{KosmiderRossier2013}%
  \BibitemOpen
  \bibfield  {author} {\bibinfo {author} {\bibfnamefont {K.}~\bibnamefont
  {Ko\ifmmode~\acute{s}\else \'{s}\fi{}mider}}, \bibinfo {author}
  {\bibfnamefont {J.~W.}\ \bibnamefont {Gonz\'alez}},\ and\ \bibinfo {author}
  {\bibfnamefont {J.}~\bibnamefont {Fern\'andez-Rossier}},\ }\bibfield  {title}
  {\bibinfo {title} {Large spin splitting in the conduction band of transition
  metal dichalcogenide monolayers},\ }\href
  {https://link.aps.org/doi/10.1103/PhysRevB.88.245436} {\bibfield  {journal}
  {\bibinfo  {journal} {Phys. Rev. B}\ }\textbf {\bibinfo {volume} {88}},\
  \bibinfo {pages} {245436} (\bibinfo {year} {2013})}\BibitemShut {NoStop}%
\bibitem [{\citenamefont {Cummings}\ \emph {et~al.}(2017)\citenamefont
  {Cummings}, \citenamefont {Garcia}, \citenamefont {Fabian},\ and\
  \citenamefont {Roche}}]{CummingsRoche2017}%
  \BibitemOpen
  \bibfield  {author} {\bibinfo {author} {\bibfnamefont {A.~W.}\ \bibnamefont
  {Cummings}}, \bibinfo {author} {\bibfnamefont {J.~H.}\ \bibnamefont
  {Garcia}}, \bibinfo {author} {\bibfnamefont {J.}~\bibnamefont {Fabian}},\
  and\ \bibinfo {author} {\bibfnamefont {S.}~\bibnamefont {Roche}},\ }\bibfield
   {title} {\bibinfo {title} {Giant spin lifetime anisotropy in graphene
  induced by proximity effects},\ }\href
  {https://link.aps.org/doi/10.1103/PhysRevLett.119.206601} {\bibfield
  {journal} {\bibinfo  {journal} {Phys. Rev. Lett.}\ }\textbf {\bibinfo
  {volume} {119}},\ \bibinfo {pages} {206601} (\bibinfo {year}
  {2017})}\BibitemShut {NoStop}%
\bibitem [{\citenamefont {Wittel}\ and\ \citenamefont
  {Manne}(1974)}]{WittelManne1974}%
  \BibitemOpen
  \bibfield  {author} {\bibinfo {author} {\bibfnamefont {K.}~\bibnamefont
  {Wittel}}\ and\ \bibinfo {author} {\bibfnamefont {R.}~\bibnamefont {Manne}},\
  }\bibfield  {title} {\bibinfo {title} {Atomic spin-orbit interaction
  parameters from spectral data for 19 elements},\ }\href
  {https://doi.org/10.1007/BF00551162} {\bibfield  {journal} {\bibinfo
  {journal} {Theoretica chimica acta}\ }\textbf {\bibinfo {volume} {33}},\
  \bibinfo {pages} {347} (\bibinfo {year} {1974})}\BibitemShut {NoStop}%
\end{thebibliography}
\end{document}